\documentclass[11pt,a4paper]{article}
\usepackage[T1]{fontenc}
\usepackage[utf8]{inputenc}
\usepackage{lmodern}
\usepackage{microtype}
\usepackage{amsmath,amssymb,amsthm,amsfonts}
\usepackage{algorithm}
\usepackage{algpseudocode}
\usepackage{tikz}

\tikzset{
  ag vertex/.style={circle,draw=black,fill=white,inner sep=1pt,
                    minimum size=6mm,font=\small,text=black},
  ag added/.style={rectangle,draw=black,fill=gray!15,inner sep=2pt,
                   minimum size=6mm,font=\small,text=black},
  ag merged/.style={ag vertex,double,double distance=1pt,
                    minimum size=7mm},
  ag match/.style={draw=black,line width=1.8pt},
  ag edge/.style={draw=black,line width=.55pt},
  ag relay/.style={draw=black,densely dotted,line width=1pt},
  ag band/.style={draw=gray!25,line width=8pt,line cap=round,
                  line join=round},
  ag label/.style={font=\scriptsize,text=black,inner sep=1pt},
  ag panel/.style={anchor=west,font=\small\bfseries,text=black},
}
\usepackage{array,booktabs}  
\usepackage[a4paper,margin=25mm]{geometry}
\usepackage{xcolor}
\definecolor{olive}{RGB}{128,128,0}
\newcommand{\corrige}[1]{#1}
\usepackage[colorlinks,linkcolor=blue!55!black,citecolor=blue!55!black]{hyperref}
\usepackage{cleveref}
\makeatletter
\def\theHALG@line{\thealgorithm.\theALG@line}
\makeatother

\newcommand{\algmin}{\textsc{MinColorMaximum}}

\newcommand{\maxtmp}{\textsc{MaxTMP}}
\newcommand{\mintmp}{\textsc{MinTMP}}
\newcommand{\algmax}{\textsc{MaxColorful}}
\newcommand{\algweight}{\textsc{MaxColorWeight}}

\newcommand{\col}{c}
\newcommand{\Cols}{\mathcal{C}}
\newcommand{\cls}[1]{V_{#1}}
\newcommand{\colset}[1]{\operatorname{col}(#1)}
\newcommand{\mult}[2]{\mu_{#1}(#2)}
\newcommand{\symdiff}{\oplus}
\newcommand{\chimax}{\chi^{*}}
\newcommand{\matchtime}{\mathcal{T}(m,n)}
\newcommand{\augtime}{\mathcal{A}(m,n)}

\theoremstyle{plain}
\newtheorem{theorem}{Theorem}[section]
\newtheorem{lemma}[theorem]{Lemma}
\newtheorem{proposition}[theorem]{Proposition}
\newtheorem{corollary}[theorem]{Corollary}
\newtheorem{observation}[theorem]{Observation}
\theoremstyle{definition}
\newtheorem{definition}[theorem]{Definition}
\newtheorem{example}[theorem]{Example}
\theoremstyle{remark}
\newtheorem{remark}[theorem]{Remark}
\crefname{observation}{Observation}{Observations}
\Crefname{observation}{Observation}{Observations}

\newcommand{\modif}[1]{#1}

\title{Tropical and maximum colorful matchings in vertex-colored graphs}
\author{%
  Johanne Cohen\thanks{LISN, Universit\'e Paris-Saclay, CNRS, Orsay, France.}
  \and
  Yannis Manoussakis\thanks{LRI, Universit\'e Paris-Saclay, CNRS, Orsay, France.}\thanks{Deceased.}
  \and
  Hong Phong Pham\footnotemark[2]
  \and
  Zsolt Tuza\thanks{HUN-REN Alfr\'ed R\'enyi Institute of Mathematics, Budapest, and
    Department of Computer Science, University of Pannonia, Veszpr\'em, Hungary.}%
}
\date{}

\begin{document}
\maketitle

\begin{abstract}
 A subgraph of a vertex-colored graph is \emph{tropical} if it contains every color. We study matchings by the number of colors they cover. A matching is \emph{maximum colorful} if no matching covers more colors. When a tropical matching exists, the maximum colorful matchings are exactly the tropical matchings. We give polynomial-time algorithms for two problems: a maximum matching that covers the most colors, and a tropical matching with the fewest edges. Both algorithms rest on a counterpart of Berge's theorem: a matching is optimal if and only if it admits no tropical augmentation, or no tropical reduction. These are chains of alternating paths linked by relays, and each one is found by a single augmenting-path search in an auxiliary graph. The same tools give a maximum colorful matching with the fewest edges, and show that all maximum matchings that are maximum colorful are connected by flips. Finally, the sets of colors that a matching can cover form a matroid. The greedy algorithm then computes a matching of maximum color weight. This paper extends the conference version~\cite{Yannis} and the PhD thesis of one of the authors~\cite{Pham18}.
\end{abstract}

\section{Introduction}\label{sec:intro}
A \emph{vertex-colored graph} is a graph with a color on each vertex. A subgraph is \emph{tropical} if it contains every color of the graph.  Finding a tropical subgraph of a given kind is usually NP-hard, even on very restricted graphs. Matchings are an exception: the problems we study here are polynomial.

Let $G^{c}$ be a vertex-colored graph. A matching is tropical if it covers at least one vertex of every color. We study two problems. The first one, \maxtmp, asks for a maximum matching that covers the most colors. If a tropical matching exists, this is a maximum tropical matching. If not, the answer tells us so, and it covers the most colors that a matching can cover. The second one, \mintmp, asks for a tropical matching with the fewest edges. Both problems were introduced in the conference version of this paper~\cite{Yannis}. The present paper is its full version, with complete proofs and explicit running times. It also contains new results on reconfiguration, on a matroid structure, and on color weights.

\textit{Related work.} Tropical subgraphs have been studied for several structures, and hardness is the rule. Angl\`es d'Auriac et al.~\cite{dAuriac18} study tropical dominating sets. The problem is NP-complete even on a path. They give bounds on the size of a solution, approximation and inapproximability results, and a fixed-parameter algorithm on interval graphs. Hudry and Lobstein~\cite{HudryLobstein22} consider rainbow sets, where every color appears exactly once, and show that rainbow domination problems have the complexity of satisfiability problems. Angl\`es d'Auriac et al.~\cite{dAuriac16} study connected tropical subgraphs. The problem is NP-hard on trees, on interval graphs, and on split graphs. It becomes tractable when the number of colors is logarithmic in $n$. Chapelle et al.~\cite{Chapelle17} give exact exponential algorithms for the same problem, in time $O^{*}(1.5359^{n})$ on general graphs and $O^{*}(1.2721^{n})$ on trees, and relate it to the Graph Motif problem. Italiano et al.\ study cliques~\cite{Italiano18cliques} and cycles~\cite{Italiano18cycles} with the most colors. For cliques, the problem is NP-hard even on classes where maximum clique is easy. Highley and Le~\cite{HighleyLe18} study tropical sets of vertex-disjoint cycles in digraphs, a model of barter exchange where every agent must trade. The problem is NP-complete and APX-hard.

Paths have received the most attention. Cohen et al.~\cite{Cohen21} study tropical paths, in two versions: the tropical path of minimum weight, and the path with the most colors. Both are NP-hard on directed acyclic graphs, on cactus graphs, and on interval graphs. The first has a fixed-parameter algorithm in general graphs. The second is polynomial on trees, block graphs, threshold graphs, bipartite chain graphs, and proper interval graphs. Fomin et al.~\cite{Fomin24} give a randomized algorithm in time $2^{k}n^{O(1)}$ for a path with at least $k$ colors between two given vertices. Bentert et al.~\cite{Bentert26} study short paths in which each color appears a prescribed number of times. The problem is NP-hard, and hard when parameterized by the number of colors, even when every color must appear equally often. It is fixed-parameter tractable in the length of the path. Dondi and Hosseinzadeh~\cite{DondiHosseinzadeh23} consider paths with the most colors in temporal graphs, and prove an inapproximability bound. The thesis of Pham~\cite{Pham18} gathers several of these results. In every case, the positive results need a restricted class of graphs or a parameter.

Matchings behave differently: the problems are polynomial on all graphs.  The algorithms of the conference version~\cite{Yannis} have since been used as a black box. Bandyapadhyay et al.~\cite{BandyapadhyayBanikBhore23} study the colorful edge cover problem, where each color class must have a prescribed number of covered vertices. They reduce it to a matching problem with a demand on each color, and this problem to tropical matching. This gives an $O(\omega n^{3})$ algorithm for $\omega$ colors. Dabas et al.~\cite{DabasKhullerRivkin25} solve the weighted version in time $O(mn+n^{2}\log n)$. So the polynomial status of tropical matchings is settled, with weights and with demands.  On the classical side, we use the matching matroid of Edmonds and Fulkerson~\cite{EdmondsFulkerson65}, the induced matroids of Rado and Perfect~\cite{Rado42,Perfect69}, and the greedy algorithm on matroids~\cite{Rado57,Edmonds71}. Reconfiguration problems were introduced in~\cite{Ito11}, and~\cite{Nishimura18} is a survey.

\textit{Our approach, and the organization of the paper.}  Our contribution is the structure behind these algorithms. We follow Berge's theorem~\cite{B57}: a matching is maximum if and only if it admits no augmenting path. \Cref{sec:prelim} introduces \emph{chains}: alternating paths connected by \emph{relays}, a relay being a covered vertex and a free vertex of the same color. \Cref{sec:max} defines \emph{tropical augmentations}, chains whose flip adds a color without changing the size. A maximum matching covers the most colors if and only if it admits no tropical augmentation (\Cref{th:MaxTM}). \Cref{sec:min} defines \emph{tropical reductions}, chains whose flip removes an edge without changing the colors. A matching has the fewest edges for its color set if and only if it admits no tropical reduction (\Cref{th:red}). In both sections, the chain is found by one augmenting-path search in an auxiliary graph. This gives the polynomial algorithms Algorithm~\algmax\ for \maxtmp\ and Algorithm~\algmin\ for \mintmp\ (\Cref{th:MaxTM2,th:MinTM2}).
 \Cref{sec:exchange} shows that Algorithm~\algmin\ also returns a maximum colorful matching with the fewest edges (\Cref{cor:colmax-min}). \Cref{sec:reconf} proves that flips connect all maximum matchings that are maximum colorful (\Cref{th:reconf}). \Cref{sec:matroid} shows that the coverable sets of colors form a matroid (\Cref{th:color-matroid}), which gives a greedy algorithm for the maximum color weight (\Cref{th:weighted2}),  (and in \Cref{sec:hall}, a variant of Hall's theorem
(\Cref{th:hall})). 
\Cref{sec:conclusion} concludes with perspectives on demands on colors.
 
\section{Preliminaries}\label{sec:prelim}

 We begin with the notation and basic tools used throughout the paper: colored matchings, flippable paths, and a counting lemma.

\subsection{Vertex-colored graphs and matchings}

Throughout the paper, $G=(V,E)$ is a finite simple undirected graph with $n=|V|$ vertices and $m=|E|$ edges. We may assume that $G$ has no isolated vertices. An isolated vertex is covered by no matching. If it is the only vertex of its color, no tropical matching exists. Otherwise it can be deleted without changing the problem. A \emph{vertex-coloring} of $G$ is a map $\col\colon V\to\Cols$, where $\Cols$ is a finite set of \emph{colors}. We denote the resulting vertex-colored graph by $G^{c}$. For each $\eta\in\Cols$, the \emph{color class} $\cls{\eta}=\col^{-1}(\eta)$ consists of the vertices of color $\eta$. For a set $S$ of vertices, we write $S_\eta=S\cap\cls{\eta}$. We write $|\Cols|$ for the number of colors, reserve $c$ for the coloring, and denote colors by Greek letters $\eta,\eta_0,\eta_1,\dots$.

A \emph{matching} of $G$ is a set $M\subseteq E$ of pairwise
vertex-disjoint edges. We write $V(M)$ for the set of vertices covered
by $M$. A vertex of $V(M)$ is \emph{$M$-covered}, and a vertex of
$V\setminus V(M)$ is \emph{$M$-free}. For an $M$-covered vertex $x$,
we denote by $M(x)$ the unique edge of $M$ containing $x$. A matching
$M$ is \emph{maximum} if no matching of $G$ has more edges.

We identify a matching $M$ with the subgraph $(V(M),M)$ of $G^{c}$.

\begin{definition}[colors of a matching, multiplicity]\label{def:colors}
The \emph{colors} of $M$ form the set
\(\colset{M}=\{\col(x) : x\in V(M)\}\).
The \emph{multiplicity} of a color $\eta$ in $M$ is the number of
$M$-covered vertices of color $\eta$:
\(\mult{M}{\eta}=|V(M)\cap\cls{\eta}|\).
A color $\eta$ is \emph{missing} in $M$ if $\mult{M}{\eta}=0$,
\emph{critical} if $\mult{M}{\eta}=1$, and \emph{redundant} if
$\mult{M}{\eta}\ge 2$.
\end{definition}

\begin{definition}[tropical matching, maximum colorful matching]\label{def:tropical}
A matching $M$ is \emph{tropical} if $\colset{M}=\Cols$: no color is
missing. We write $\chimax$ for the largest number of colors that a
matching can cover. A matching is  \emph{maximum colorful} if $|\colset{M}|=\chimax$. A tropical
matching exists if and only if $\chimax=|\Cols|$, and in this case
the tropical matchings are exactly the maximum colorful matchings.
\end{definition}

An $M$-covered vertex $x$ is \emph{critical} in $M$ if its color is
critical in $M$, that is, if $x$ is the only $M$-covered vertex of
color $\col(x)$.

\subsection{Alternating paths and flips}

For a path $P$ of $G$, we write $V(P)$ and $E(P)$ for its sets of
vertices and edges. A path $P$ is \emph{$M$-alternating} if its edges
alternate between $M$ and $E\setminus M$~\cite{B57,LovaszPlummer86}.
Replacing $M$ by $M\symdiff E(P)$, where
$X\symdiff Y=(X\setminus Y)\cup(Y\setminus X)$ denotes the symmetric
difference of two sets, is a classical operation in matching theory.
When both endpoints of $P$ are $M$-free, $P$ is an
\emph{$M$-augmenting path}, and the operation increases the size of the
matching by one. Iterating Berge's theorem gives the following lemma.

\begin{lemma}\label{lem:berge-ext}
For every matching $M$ of $G$, there is a maximum matching $M^*$ of $G$
with $V(M)\subseteq V(M^*)$. In particular,
$\colset{M}\subseteq\colset{M^*}$.
\end{lemma}

\begin{proof}
If $M$ is not maximum, it has an augmenting path $P$ \cite{B57}. The
matching $M\symdiff E(P)$ has one more edge, and it covers $V(M)$ and
the two endpoints of $P$. Iterating, we reach a maximum matching that
covers $V(M)$.
\end{proof}

 Applied to a maximum colorful matching, the lemma gives
a maximum matching that is also maximum colorful.

 More generally, in what follows we replace $M$ by $M\symdiff E(P)$ for any path $P$ for which this is still a matching. Although this condition is used implicitly in many arguments, to the best of our knowledge it has  no established name. Since we use this condition throughout the paper, we give it  one.

\begin{definition}[flippable path]\label{def:flippable}
An $M$-alternating path $P$ with at least one edge is
\emph{$M$-flippable} if $M\symdiff E(P)$ is a matching.
\emph{Flipping} $P$ means replacing $M$ by $M\symdiff E(P)$.
\end{definition}

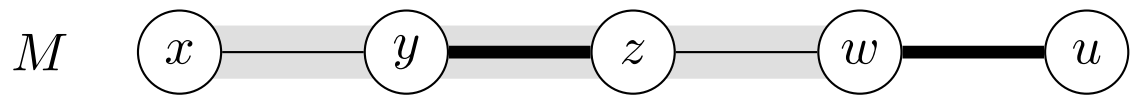
\begin{figure}[htb]
\centering
\begin{tikzpicture}[
  vertex/.style={circle, draw, fill=white, inner sep=0pt,
                 minimum size=5.5mm, font=\small},
  medge/.style={line width=2.4pt},
  oedge/.style={thin},
  pathbg/.style={line width=10pt, gray!25, line cap=round},
  rowlabel/.style={anchor=east, font=\small, align=right},
]
\begin{scope}[name prefix=a-]
  \draw[pathbg] (0,0) -- (4.5,0);
  \foreach \v/\X in {x/0, y/1.5, z/3, w/4.5, u/6}
    \node[vertex] (\v) at (\X,0) {$\v$};
  \draw[oedge] (x) -- (y);  \draw[medge] (y) -- (z);
  \draw[oedge] (z) -- (w);  \draw[medge] (w) -- (u);
  \node[rowlabel] at (-0.6,0) {$M$};
\end{scope}
\end{tikzpicture}
\caption{The path $x,y,z,w$ (gray band) of \Cref{ex:not-flippable} is
$M$-alternating but not $M$-flippable. Thick edges belong to $M$, thin
edges do not.}
\label{fig:not-flippable}
\end{figure}

\begin{example}\label{ex:not-flippable}
Let $G$ be the path $x,y,z,w,u$ and $M=\{yz,wu\}$
(\Cref{fig:not-flippable}). The path $P=x,y,z,w$ is $M$-alternating,
since $xy\notin M$, $yz\in M$ and $zw\notin M$. Its endpoint $x$ is
$M$-free, but its endpoint $w$ is covered by $wu$, which is not an edge
of $P$. Flipping $P$ would give $M\symdiff E(P)=\{xy,zw,wu\}$, in which
$w$ is covered twice. Hence $P$ is not $M$-flippable.
\end{example}

Only the endpoints can affect whether the flip produces a matching. Let $x$ be an endpoint of $P$, and let $e$ be its incident edge on $P$. Each internal vertex has one incident edge in $M$ and one outside $M$; the flip exchanges these edges, so the vertex remains covered exactly once. An endpoint becomes covered twice precisely when $e\notin M$, $x$ is $M$-covered, and $M(x)\notin E(P)$. The following observation states the resulting criterion.

\begin{observation}\label{obs:flippable}
Let $P$ be an $M$-alternating path with at least one edge. Then $P$
is $M$-flippable if and only if each endpoint $v$ of $P$ is $M$-free
or satisfies $M(v)\in E(P)$.
\end{observation}

Hence, when we flip an $M$-flippable path $P$, only its endpoints change status. An $M$-free endpoint becomes covered. An $M$-covered endpoint becomes free. All other vertices keep their status.

 An \emph{$M$-alternating cycle} $C$ is a cycle whose edges alternate between $M$ and $E\setminus M$. Flipping $C$ means replacing $M$ by $M\symdiff E(C)$. The result is a matching. It has the same number of edges as $M$, and it covers the same vertices.

\subsection{Symmetric difference of two matchings}

Our flippable paths always come from comparing two matchings. Let
$M$ and $N$ be matchings of $G$, and let
$G_{\symdiff}=(V,M\symdiff N)$. Every component of $G_{\symdiff}$
with at least one edge is an even cycle or a path, and its edges
alternate between $M$ and $N$~\cite{B57}.

\begin{lemma}\label{lem:symdiff}
Let $M$ and $N$ be matchings of $G$, and let
$G_{\symdiff}=(V,M\symdiff N)$. Every path component $P$ of
$G_{\symdiff}$ is both $M$-flippable and $N$-flippable.
\end{lemma}

\begin{proof}
Let $x$ be an endpoint of $P$, and $e$ its edge in $P$. Assume
$e\in M\setminus N$ (the other case is symmetric). Since $x$ has
degree $1$ in $G_{\symdiff}$, $x$ is $N$-free, and $M(x)=e\in E(P)$.
So $x$ satisfies the condition of \Cref{obs:flippable} for both $M$
and $N$.
\end{proof}

The proof shows that each endpoint of $P$ is covered by exactly one of
$M$ and $N$, namely the one that contains its edge on $P$.

\subsection{Relays and flippable chains}

Our algorithms flip several paths at once. Relays link them, and
chains organize them.

\begin{definition}[relay]\label{def:relay}
Let $P$ and $Q$ be two vertex-disjoint $M$-flippable paths. A
\emph{relay} is a pair of vertices $\{u,v\}$ where $u$ is an endpoint of $P$, $v$
is an endpoint of $Q$, $\col(u)=\col(v)$, and
one of $u$ and $v$ is $M$-covered while the other is $M$-free. The color
$\col(u)=\col(v)$ is the  color of the relay.
\end{definition}

Let $\{u,v\}$ be a relay, with $u$ $M$-covered and $v$ $M$-free. After flipping $P$ and $Q$, the vertex $u$ is free
and the vertex $v$ is covered by the new matching. So the color
$\col(u)=\col(v)$ keeps its multiplicity: it is now covered at $v$
instead of $u$.

\begin{definition}[flippable chain]\label{def:chain}
An \emph{$M$-flippable chain} of order $k$ is a sequence
$\mathcal{P}=(P_1,\dots,P_k)$ of pairwise vertex-disjoint
$M$-flippable paths, each $P_i$ being oriented from its \emph{start}
$x_i$ to its \emph{end} $x'_i$, such that $\{x'_i,x_{i+1}\}$ is a
relay for all $1\le i<k$. The \emph{terminals} of the chain are
$x_1$ and $x'_k$. Flipping the chain means flipping all its paths.
We write $M\symdiff\mathcal{P}$ for the resulting matching, that is, $M\symdiff E(P_1)\symdiff\cdots\symdiff E(P_k)$. We write $V(\mathcal{P})$ and $E(\mathcal{P})$ for the sets of vertices and edges of the paths of $\mathcal{P}$.
\end{definition}

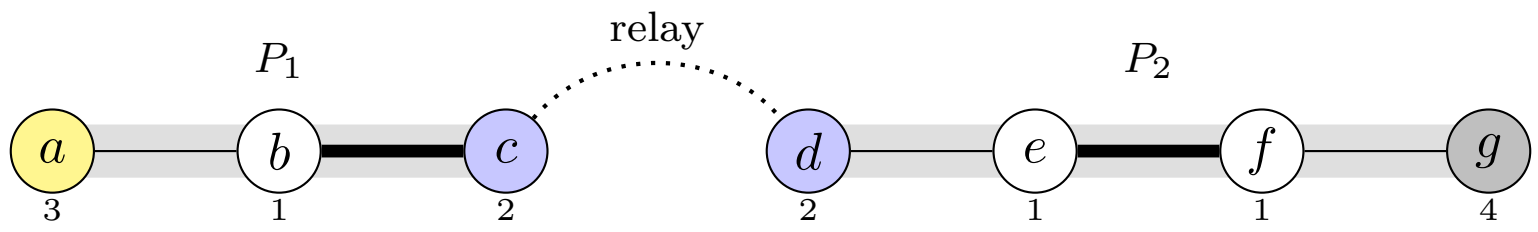
\begin{figure}[htb]
\centering
\begin{tikzpicture}[
  vertex/.style={circle, draw, inner sep=0pt, minimum size=5.5mm,
                 font=\small},
  medge/.style={line width=2.4pt},
  oedge/.style={thin},
  pathbg/.style={line width=10pt, gray!25, line cap=round},
  cnum/.style={font=\tiny, inner sep=1pt},
  tag/.style={font=\scriptsize},
]
  \draw[pathbg] (0,0) -- (3,0);
  \draw[pathbg] (5,0) -- (9.5,0);
  \node[vertex, fill=yellow!55, label={[cnum]below:3}] (a) at (0,0) {$a$};
  \node[vertex, fill=white,     label={[cnum]below:1}] (b) at (1.5,0) {$b$};
  \node[vertex, fill=blue!22,   label={[cnum]below:2}] (c) at (3,0) {$c$};
  \node[vertex, fill=blue!22,   label={[cnum]below:2}] (d) at (5,0) {$d$};
  \node[vertex, fill=white,     label={[cnum]below:1}] (e) at (6.5,0) {$e$};
  \node[vertex, fill=white,     label={[cnum]below:1}] (f) at (8,0) {$f$};
  \node[vertex, fill=gray!50,   label={[cnum]below:4}] (g) at (9.5,0) {$g$};
  \draw[oedge] (a) -- (b); \draw[medge] (b) -- (c);
  \draw[oedge] (d) -- (e); \draw[medge] (e) -- (f); \draw[oedge] (f) -- (g);
  \draw[dotted, thick] (c) to[out=50,in=130] (d);
  \node[tag] at (4,0.8) {relay};
  \node[tag] at (1.5,0.6) {$P_1$};
  \node[tag] at (7.25,0.6) {$P_2$};
\end{tikzpicture}
\caption{A chain $(P_1,P_2)$ with $P_1=a,b,c$ and $P_2=d,e,f,g$. The
color of a vertex is shown by its fill and by the number below it.
Thick edges belong to $M=\{bc,ef\}$. The vertices $c$ ($M$-covered endpoint) and $d$ ($M$-free endpoint) have the same color~2, so
$\{c,d\}$ is a relay. The terminals are $a$ and $g$, both free.
Flipping the chain adds one edge and the colors 3 and~4, and keeps
the multiplicity of color~2.}
\label{fig:chain}
\end{figure}

 By the counting lemma below, flipping a chain can change the matching size and color multiplicities only at its two terminals, $x_1$ and $x'_k$. 
 See \Cref{fig:chain}.

The next lemma is the single tool behind all our flips.

\begin{lemma}[counting lemma]\label{lem:counting}
Let $M$ be a matching of $G^{c}$ and let $\mathcal{P}$ be an
$M$-flippable chain, with set $A$ of $M$-free endpoints and set $B$ of $M$-covered endpoints. Then $N=M\symdiff\mathcal{P}$ is a matching with
\(
V(N) \;=\; \bigl(V(M)\setminus B\bigr)\cup A\),
and consequently
\[
|N| \;=\; |M| + \frac{|A|-|B|}{2}
\qquad\text{and}\qquad
\mult{N}{\eta} \;=\; \mult{M}{\eta} - |B_\eta| + |A_\eta|
\quad\text{for every color }\eta\in\Cols .
\]
\end{lemma}

\begin{proof}
Flipping a path $P_i$ only changes edges with both endpoints in
$V(P_i)$. As the paths are vertex-disjoint, each $P_i$ is still flippable after
$P_1,\dots,P_{i-1}$ have been flipped.  So, $N$ is obtained by flipping them one
at a time. Each flip yields a matching and only changes the status of
the endpoints of $P_i$ (see after \Cref{obs:flippable}). This proves
the formula for $V(N)$.

Since $A\cap V(M)=\emptyset$ and $B\subseteq V(M)$, we get
$|V(N)|=|V(M)|-|B|+|A|$, hence the formula for $|N|$. Applying the
same count to $\cls{\eta}$ gives the formula for $\mult{N}{\eta}$.
\end{proof}

Flipping an alternating cycle changes neither the size of $M$ nor the set $V(M)$. Our algorithms instead flip chains of paths. We use two types: one preserves the matching size and adds a color (\Cref{sec:max}); the other preserves the color set and removes one edge (\Cref{sec:min}).

We end with the case of two maximum matchings, which we use in
\Cref{sec:max} and in \Cref{sec:reconf}.

\begin{lemma}[two maximum matchings]\label{lem:two-max}
Let $M$ and $N$ be two maximum matchings of $G$, and let
$G_{\symdiff}=(V,M\symdiff N)$.
\begin{enumerate}
\item Every path of $G_{\symdiff}$ has one endpoint covered by
  $N$ and free for $M$, its \emph{start}, and one endpoint
  covered by $M$ and free for $N$, its \emph{end}.
\item Every vertex covered by $N$ and free for $M$ is the start
  of a path of $G_{\symdiff}$.
\item Let $P_1,\dots,P_k$ be distinct paths of $G_{\symdiff}$
  such that, for $1\le i<k$, the end of $P_i$ and the start of
  $P_{i+1}$ have the same color. Then $(P_1,\dots,P_k)$ is an
  $M$-flippable chain, and flipping it keeps the size of $M$.
\end{enumerate}
\end{lemma}

\begin{proof}
(1) Let $P$ be a path of $G_{\symdiff}$. Its edges alternate
between $M$ and $N$. Since $M$ and $N$ are both maximum, $P$ has
as many edges in $M$ as in $N$: otherwise, it would be an
augmenting path for the matching with fewer edges on it,
contradicting Berge's theorem~\cite{B57}. So the edge of $P$ at
one endpoint is in $N$, and the edge at the other endpoint is in
$M$. The first endpoint is the start of $P$, and the second is
its end. We orient $P$ from its start to its end. As noted after
\Cref{lem:symdiff}, each endpoint is covered by exactly one of
$M$ and $N$, the one that contains its edge on $P$. So the start
is $N$-covered and $M$-free, and the end is $M$-covered and
$N$-free.

(2) Let $u$ be a vertex that is $N$-covered and $M$-free. Its
$N$-edge is not in $M$, so it is an edge of $G_{\symdiff}$. It
is its only edge there, since $u$ is $M$-free. So $u$ has degree
$1$ in $G_{\symdiff}$: it is an endpoint of a path. Its edge on
this path is in $N$, so $u$ is the start of the path.

(3) Distinct paths of $G_{\symdiff}$ are vertex-disjoint, and
they are $M$-flippable by \Cref{lem:symdiff}. For $1\le i<k$,
the end of $P_i$ is $M$-covered, the start of $P_{i+1}$ is
$M$-free, and they have the same color: this pair is a relay.
So $(P_1,\dots,P_k)$ is an $M$-flippable chain in which every
start is $M$-free and every end is $M$-covered. It has $k$
$M$-free endpoints and $k$ $M$-covered endpoints. By
\Cref{lem:counting}, flipping it keeps the size of $M$.
\end{proof}

\subsection{Complexity conventions}\label{sec:complexity}

 We close this section with two conventions on running
times.

Let $\matchtime$ denote the time needed to compute a maximum matching in a graph with $n$ vertices and $m$ edges. Let $\augtime$ denote the time needed either to find an augmenting path for a given matching or to report that none exists. Both functions are polynomial: $\matchtime=O(\sqrt{n}\,m)$~\cite{MicaliVazirani80} and $\augtime=O(m\,\alpha(m,n))$, where $\alpha$ is an inverse of Ackermann's function~\cite{Gabow76,Tarjan83}. We use these functions only as black boxes on auxiliary graphs with $O(n)$ vertices and $O(m+n)$ edges. Since $G$ has no isolated vertices, $n\le 2m$, so these graphs have $O(m)$ edges. Thus the bounds expressed by $\matchtime$ and $\augtime$ retain the same asymptotic order on the auxiliary graphs. 
 In the statements below, we also give the explicit bounds
obtained from these two estimates. The functions $\matchtime$ and
$\augtime$ remain meaningful on their own: on special graph classes
they can be smaller, and the bounds improve accordingly. In
bipartite graphs, for instance, an augmenting path can be found in
time $O(m)$.
\section{Maximum tropical matchings}\label{sec:max}
By \Cref{lem:berge-ext}, to cover as many colors as possible, it
suffices to consider maximum matchings.

\subsection{Tropical augmentations}

Our basic operation is a \emph{tropical augmentation} defined in~\cite{Yannis}: flipping the chain preserves the matching size and adds one covered color.

\begin{definition}[tropical augmentation]\label{def:k-tropical-aug}
Let $M$ be a matching of $G^{c}$. A \emph{tropical augmentation} of
$M$ of order $k\ge1$ is an $M$-flippable chain
$\mathcal{P}=(P_1,\dots,P_k)$ in which each start $x_i$ is an $M$-free endpoint and each end $x'_i$ is an $M$-covered endpoint, and such that
\begin{enumerate}
\item the color $\col(x_1)$ is missing in $M$,
\item the color of each relay $\{x'_i,x_{i+1}\}$, for $1\le i<k$, is
  critical in $M$, and
\item the color $\col(x'_k)$ is redundant in $M$.
\end{enumerate}
\end{definition}

\begin{figure}[htb]
\centering
\begin{tikzpicture}[
  vertex/.style={circle, draw, inner sep=0pt, minimum size=5.5mm,
                 font=\small},
  medge/.style={line width=2.4pt},
  oedge/.style={thin},
  pathbg/.style={line width=10pt, gray!25, line cap=round},
  cnum/.style={font=\tiny, inner sep=1pt},
  tag/.style={font=\scriptsize},
]
  \draw[pathbg] (0,0) -- (3,0);
  \draw[pathbg] (5,0) -- (8,0);
  \node[vertex, fill=gray!50,   label={[cnum]below:4}] (a) at (0,0) {$a$};
  \node[vertex, fill=blue!22,   label={[cnum]below:2}] (b) at (1.5,0) {$b$};
  \node[vertex, fill=yellow!55, label={[cnum]below:3}] (c) at (3,0) {$c$};
  \node[vertex, fill=yellow!55, label={[cnum]below:3}] (d) at (5,0) {$d$};
  \node[vertex, fill=blue!22,   label={[cnum]below:2}] (e) at (6.5,0) {$e$};
  \node[vertex, fill=white,     label={[cnum]below:1}] (f) at (8,0) {$f$};
  \node[vertex, fill=white,     label={[cnum]below:1}] (g) at (10,0) {$g$};
  \node[vertex, fill=blue!22,   label={[cnum]below:2}] (h) at (11.5,0) {$h$};
  \draw[oedge] (a) -- (b); \draw[medge] (b) -- (c);
  \draw[oedge] (d) -- (e); \draw[medge] (e) -- (f);
  \draw[medge] (g) -- (h);
  \draw[dotted, thick] (c) to[out=50,in=130] (d);
  \node[tag] at (4,0.8) {relay};
  \node[tag] at (0.1,0.7) {color 4 is missing};
  \node[tag] at (9,0.7) {color 1 is redundant};
  \node[tag] at (1.7,0.6) {$P_1$};
  \node[tag] at (6.5,0.6) {$P_2$};
\end{tikzpicture}
\caption{The matching $M=\{bc,ef,gh\}$ covers the colors $1$, $2$
and $3$; the color $4$ is missing, the color $3$ is critical (only
$c$), and the color $1$ is redundant ($f$ and $g$). The chain
$(P_1,P_2)$ with $P_1=a,b,c$, $P_2=d,e,f$ and the relay $\{c,d\}$ is
a tropical augmentation. After the flip, $N=\{ab,de,gh\}$ has the
same size and covers all four colors.}
\label{fig:aug-ex}
\end{figure}
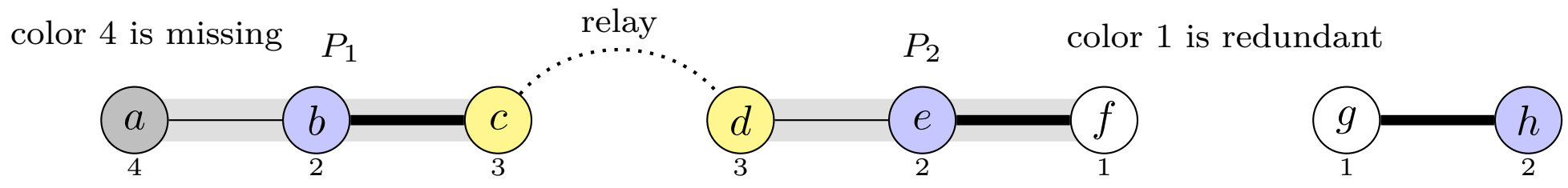

Exactly one of the two edges of $P_i$ at its endpoints is in $M$, the one at $x'_i$. The relay colors are pairwise distinct: a critical color has a single
$M$-covered vertex, and the ends $x'_1,\dots,x'_{k-1}$ are distinct.
See \Cref{fig:aug-ex}.

\begin{lemma}\label{lem:flip-aug}\cite{Yannis}
If $\mathcal{P}$ is a tropical augmentation of $M$, then
$N=M\symdiff\mathcal{P}$ is a matching of $G^{c}$ with $|N|=|M|$ and
$\colset{N}=\colset{M}\cup\{\col(x_1)\}$. In particular,
$|\colset{N}|=|\colset{M}|+1$.
\end{lemma}

\begin{proof}
By \Cref{lem:counting}, $N$ is a matching. Each $P_i$ has one $M$-free endpoint $x_i$ and one $M$-covered endpoint $x'_i$, so $|A|=|B|=k$ and
$|N|=|M|$. Fix a color $\eta$. Each relay of color $\eta$ puts one
vertex in $A_\eta$ and one vertex in $B_\eta$, so its contribution
to $\mult{N}{\eta}=\mult{M}{\eta}-|B_\eta|+|A_\eta|$ is zero. Hence
only the colors of the terminals $x_1$ and $x'_k$ change
multiplicity. These two colors are distinct: one is missing and the other is
redundant in $M$. The color $\col(x_1)$ goes from multiplicity $0$
in $M$ to $1$ in $N$: it is the new color of $N$. The color
$\col(x'_k)$ goes from $\mult{M}{\col(x'_k)}\ge 2$ in $M$ to
$\mult{M}{\col(x'_k)}-1\ge 1$ in $N$: it stays covered in $N$. Hence
$\colset{N}=\colset{M}\cup\{\col(x_1)\}$.
\end{proof}

\subsection{The augmentation theorem}

 The proof relies on choosing the  matching $N$ carefully.

\begin{lemma}[choice of $N$ for augmentations]\label{lem:choice-aug}
Let $M$ be a maximum matching of $G^{c}$, and assume that some
maximum matching of $G^{c}$ has more colors than $M$. Among these
matchings, let $N$ be one with $|M\cap N|$ maximum, and let
$G_{\symdiff}=(V,M\symdiff N)$. Then:
\begin{enumerate}
\item every component of $G_{\symdiff}$ with at least one edge is a
  path, with as many edges in $M$ as in $N$;
\item the colors of the two endpoints of every path component of
  $G_{\symdiff}$ are covered by $N$.
\end{enumerate}
\end{lemma}

\begin{proof}
 
(1) The components of $G_{\symdiff}$ with at least one edge are even
cycles and paths whose edges alternate between $M$ and $N$. Suppose
that some component is a cycle $C$, and let $N'=N\symdiff E(C)$. It
is a matching with $V(N')=V(N)$, so it is maximum with
$\colset{N'}=\colset{N}$, and it contains the edges of $M$ on $C$:
$|M\cap N'|>|M\cap N|$. This contradicts the choice of $N$. So
every component with at least one edge is a path. Since $M$ and $N$
are both maximum, no path has more edges in one of them than in the
other: otherwise, it would be an augmenting path for the matching
with fewer edges on it, contradicting Berge's theorem~\cite{B57}.

(2) Let $P$ be a path component, with endpoints $x$ and $y$. By (1),
the edge of $P$ at one endpoint is in $N$ and the edge at the other is in $M$; say the edge at $x$ is in $N$ and the edge at $y$ is in $M$. So, $x$ is covered by $N$ and not by $M$, and $y$
is covered by $M$ and not by $N$. So $\col(x)$ is covered by $N$,
at $x$ itself. Assume now that $\col(y)\notin\colset{N}$, and let
$N'=N\symdiff E(P)$. It is a maximum matching, equal to $M$ on $P$
and to $N$ elsewhere. In particular, $|M\cap N'|>|M\cap N|$. It
covers $y$ instead of $x$. So $\colset{N'}$ contains the color $\col(y)$, and it contains every color of $\colset{N}$ except possibly $\col(x)$. Since
$\col(y)\notin\colset{N}$, this gives
$|\colset{N'}|\ge|\colset{N}|>|\colset{M}|$. This contradicts the
choice of $N$.\end{proof}
 
 The next theorem characterizes when a tropical augmentation exists and provides the main structural result behind our algorithm for \maxtmp.

\begin{theorem}\label{th:MaxTM}
Let $M$ be a maximum matching of $G^{c}$. Then $M$ admits a tropical
augmentation if and only if there exists a maximum matching $N$ of
$G^{c}$ with $|\colset{N}| > |\colset{M}|$.
\end{theorem}

\begin{proof}
If $\mathcal{P}$ is a tropical augmentation of $M$, then by
\Cref{lem:flip-aug} the matching $N=M\symdiff\mathcal{P}$ has the same
size as $M$, so it is maximum, and it has one more color.

Conversely,  we assume that some maximum matching has
more colors than $M$. {We choose} $N$ as in
\Cref{lem:choice-aug}, and let $\mathcal{Q}$ be the set of path
components of $G_{\symdiff}=(V,M\symdiff N)$. {Since $M$
and $N$ are both maximum, \Cref{lem:two-max} applies. Every
path of $\mathcal{Q}$ has a start, which is $N$-covered and
$M$-free, and an end, which is $M$-covered and $N$-free. We
orient it from its start to its end.} However, by
\Cref{lem:choice-aug}(2), the color of the end is covered by
$N$, at another vertex.  {Moreover, every vertex that is
$N$-covered and $M$-free is the start of a path of
$\mathcal{Q}$.

\smallskip
\noindent\textit{Step 1: we build a longest admissible
sequence.}
Since $|\colset{N}|>|\colset{M}|$, there is a color
$\eta_0\in\colset{N}\setminus\colset{M}$. An $N$-covered vertex
of color $\eta_0$ is $M$-free, so it is the start of a path
$P_1\in\mathcal{Q}$. We consider a sequence $(P_1,\dots,P_k)$
of distinct paths of $\mathcal{Q}$ beginning with this path
$P_1$. We call it \emph{admissible} if, for $1\le i<k$, the end
of $P_i$ and the start of $P_{i+1}$ have the same color,
critical in $M$.  By \Cref{lem:two-max}(3), an admissible
sequence is an $M$-flippable chain in which every start is
$M$-free and every end is $M$-covered. It satisfies conditions
(1) and (2) of \Cref{def:k-tropical-aug}: the color $\eta_0$ of
its first start is missing in $M$, and its relay colors are
critical in $M$.

The sequence $(P_1)$ is admissible and $\mathcal{Q}$ is finite,
so we can choose an admissible sequence $(P_1,\dots,P_k)$ with
$k$ maximum. Let $x'_k$ be the end of $P_k$, and let $\eta_k$ be
its color. Since $x'_k$ is $M$-covered, $\eta_k$ is critical or
redundant in $M$.

\smallskip
\noindent\textit{Step 2: we prove that $\eta_k$ is
redundant in $M$.}
 This is condition (3) of \Cref{def:k-tropical-aug}. We
suppose, for contradiction,} that $\eta_k$ is critical in $M$.
Then $x'_k$ is the only $M$-covered vertex of color $\eta_k$. By
\Cref{lem:choice-aug}(2), some vertex $y\neq x'_k$ of color
$\eta_k$ is $N$-covered. Since $\eta_k$ is critical, $y$ is
$M$-free. Hence $y$ is the start of a path $P\in\mathcal{Q}$.
This path is none of $P_1,\dots,P_k$. Otherwise $y$ would be the
start of some $P_j$. We have $j\neq1$, since the start of $P_1$
has the missing color $\eta_0$, while $\eta_k$ is covered by
$M$. And if $j\ge2$, the end of $P_{j-1}$ is an $M$-covered
vertex of color $\eta_k$. It differs from $x'_k$ because
$j-1<k$, which contradicts the criticality of $\eta_k$. So
$(P_1,\dots,P_k,P)$ is admissible, which contradicts the
maximality of $k$.

 Hence $\eta_k$ is redundant in $M$, and
$(P_1,\dots,P_k)$ is a tropical augmentation of $M$.
\end{proof}

 By \Cref{th:MaxTM} and \Cref{lem:berge-ext}, a maximum
matching without tropical augmentation is maximum colorful.

\subsection{Computing a maximum colorful matching: Algorithm~\algmax}

We now turn \Cref{th:MaxTM} into an algorithm. We start from a maximum
matching of $G$ and modify it to cover one more color, as long as there exist
maximum matching having more colors. Each step builds a tropical
augmentation.  We find it by one
augmenting-path search in an auxiliary graph. We first state the algorithm
and its running time, using this search as a black box. The quotient
graph is built in \Cref{sec:find-aug}.

\begin{algorithm}[htb]
\caption{\algmax\ (maximum matching that is maximum colorful)}\label{alg:MaxTM}
\begin{algorithmic}[1]
\State $M \gets$ a maximum matching of $G$.
\While{$M$ admits a tropical augmentation $\mathcal{P}$
  (\Cref{pr:MaxTM} below)}
  \State $M \gets M\symdiff\mathcal{P}$.
\EndWhile
\State \Return $M$. \textit{($M$ is tropical if and only if
  $|\colset{M}|=|\Cols|$)}.
\end{algorithmic}
\end{algorithm}

\begin{theorem}\label{th:MaxTM2}
 Algorithm~\algmax\ (\Cref{alg:MaxTM}) computes a maximum matching  of $G^{c}$ that is maximum colorful
in time
$O(\matchtime+|\Cols|\,\augtime) =O\bigl(\sqrt{n}\,m+|\Cols|\,m\,\alpha(m,n)\bigr)$.
\end{theorem}

\begin{proof}
Line~1 takes time $O(\matchtime)$. By \Cref{pr:MaxTM}, the current matching stays maximum and
gains at least one color at each iteration, so there are at most
$|\Cols|$ iterations. Each iteration builds $G'$ and $K$ in time
$O(n+m)$, and searches one augmenting path, in time $O(\augtime)$.

When the loop stops, $M$ is maximum and admits no tropical
augmentation. By \Cref{th:MaxTM}, no maximum matching has more colors
than $M$. By \Cref{lem:berge-ext}, no matching at all has more colors
than $M$. So $M$ is maximum colorful.

\end{proof}

 In particular, Algorithm~\algmax\ solves \maxtmp: it returns a maximum tropical matching if one exists.

\subsection{Finding augmentations: the quotient graph}\label{sec:find-aug}

We now construct the auxiliary graph used at each iteration of
Algorithm~\algmax.

Let $M$ be a maximum matching of $G^{c}$. For a critical color $\eta$,
let $x_\eta$ denote the unique $M$-covered vertex of color $\eta$. The
\emph{quotient graph} $G'=G'(G^{c},M)$ is obtained from $G^{c}$ and
$M$ in three steps.
\begin{enumerate}
\item Delete every $M$-free vertex whose color is redundant in $M$.
\item For every critical color $\eta$, merge the vertices of color
  $\eta$ into a single vertex $v_\eta$, deleting loops and keeping one
  copy of each edge.
\item Add a new vertex $z$, adjacent to every $M$-covered vertex whose
  color is redundant in $M$.
\end{enumerate}
The graph $G'$ has at most $n+1$ vertices and at most $m+n$ edges.

We call the vertices $v_\eta$ the \emph{merged vertices}. A vertex of
$G$ that is not deleted has an \emph{image} in $G'$. It is $v_\eta$ if
the color $\eta$ of the vertex is critical, and the vertex itself
otherwise. An edge of $G$ whose endpoints have distinct images has an
image too: the edge of $G'$ that joins them. Every edge of $G'$ that
is not at $z$ is the image of an edge of $G$.

Let $M'$ be the image of $M$ in $G'$. An edge of $M$ with no endpoint
of critical color is unchanged: it is also an edge of $M'$. The other
edges of $M$ are the edges $M(x_\eta)$, where $\eta$ is a critical
color. In $M'$, such an edge joins $v_\eta$ to the image of its other
endpoint. Since two $M$-covered vertices are never merged, $M'$ is a
matching of $G'$ with $|M'|=|M|$, and each $v_\eta$ is covered exactly
once. The $M'$-free vertices of $G'$ are $z$ and the $M$-free vertices
whose color is missing in $M$.

 Step~2 merges the two vertices of
each relay, so the paths of the chain follow one another in $G'$.
Step~3 lets the path continue to $z$ after an end of redundant color.
Step~1 removes vertices that belong to no tropical augmentation.

\begin{proposition}\label{pr:MaxTM}
Let $M$ be a maximum matching of $G^{c}$, let $G'=G'(G^{c},M)$ be the
quotient graph, and let $M'$ be the image of $M$ in $G'$. Then $M$
admits a tropical augmentation if and only if $G'$ has an
$M'$-augmenting path. More precisely:
\begin{enumerate}
\item Let $(P_1,\dots,P_k)$ be a tropical augmentation of $M$. In
  $G'$, the images of $P_1,\dots,P_k$ follow one another and form a
  single path. Adding $z$ at its end gives an $M'$-augmenting path;
\item every $M'$-augmenting path of $G'$ is obtained in this way from
  a tropical augmentation of $M$, which can be computed from the path
  in time $O(n+m)$.
\end{enumerate}
\end{proposition}

\begin{proof}
(1) Assume first that $M$ has a tropical augmentation
$\mathcal{P}=(P_1,\dots,P_k)$, with starts $x_i$ and ends $x'_i$.
Each $P_i$ is $M$-alternating. Its start $x_i$ is $M$-free. Its end
$x'_i$ is $M$-covered, and its last edge is $M(x'_i)$. Step~2 merges
the two vertices $x'_i$ and $x_{i+1}$ of each relay. So in $G'$ the
end of $P_i$ is the start of $P_{i+1}$. Hence $P_1,\dots,P_k$ form a
single $M'$-alternating path from $x_1$ to $x'_k$.

The start $x_1$ is $M$-free, and its color is missing in $M$. So it
is not deleted in step~1, and it is $M'$-free. The end $x'_k$ is
$M$-covered, and its color is redundant in $M$. So it is adjacent to
$z$. The edge $x'_kz$ is not in $M'$, and $z$ is $M'$-free. Adding
$z$ at the end gives an $M'$-augmenting path from $x_1$ to $z$.

(2) Conversely, let $W$ be an $M'$-augmenting path of $G'$. Its two
endpoints are $M'$-free, and its internal vertices are $M'$-covered.
At most one endpoint is $z$. Hence one endpoint of $W$ is an $M$-free
vertex $x$ of $G^{c}$ whose color is missing in $M$. The other
endpoint $x'$ is either $z$ or another such vertex. We orient $W$
from $x$ to $x'$, and we prove below that $x'=z$.

We transform $W$ into paths of $G$. If the last vertex of $W$ is $z$,
we first remove it, with its edge. Every other vertex of $W$ that is
not merged is a vertex of $G$, and we keep it. Every remaining edge of
$W$ is the image of an edge of $G$. We choose one such edge of $G$ for
each edge of $W$. For an edge of $M'$, we choose the edge of $M$.

 We consider a merged vertex $v_\eta$ of $W$. It is an
internal vertex of $W$. Its edge of $M'$ is the edge $M(x_\eta)$ of
$G$, at the $M$-covered vertex $x_\eta$. Its other edge on $W$ is an
edge of $G$ at some vertex $y$ of color $\eta$. Either $y=x_\eta$,
and the path goes through $x_\eta$. Or $y$ is an $M$-free vertex, and
$v_\eta$ corresponds to the two distinct vertices $x_\eta$ and $y$ of
$G$. In this second case, the path is cut at $v_\eta$: one part ends
at $x_\eta$, and the next part starts at $y$, or the reverse. The
part that contains $x_\eta$ also contains the edge $M(x_\eta)$.

We get vertex-disjoint $M$-alternating paths $Q_1,\dots,Q_k$ of $G$,
in the order of $W$. None of them has two $M$-free endpoints.
Otherwise it would be an $M$-augmenting path, which contradicts the
maximality of $M$.

The path $Q_1$ starts at $x$, which is $M$-free. So it ends at an
$M$-covered vertex. If $k>1$, this end is at a cut. Since it is
$M$-covered, it is the vertex $x_{\eta_1}$ for some critical color
$\eta_1$, and $Q_2$ starts at an $M$-free vertex of color $\eta_1$.
So $Q_2$ ends at an $M$-covered vertex, and so on. Hence every $Q_i$
starts at an $M$-free vertex and ends at an $M$-covered vertex. For
$i<k$, the end of $Q_i$ and the start of $Q_{i+1}$ form a relay,
and its color is critical in $M$.

The end of $Q_k$ is $M$-covered, so it is not $x'$. Hence $x'=z$,
and $Q_k$ ends at the neighbor $w$ of $z$ on $W$. The color of $w$ is
redundant in $M$. The edge $wz$ is not in $M'$, so the last edge of
$Q_k$ is $M(w)$.

Each $Q_i$ has an $M$-free start, and the edge of $M$ at its end is
its last edge. So it is $M$-flippable by \Cref{obs:flippable}. The
color of $x$ is missing in $M$, the relay colors are critical, and
the color of $w$ is redundant. So $(Q_1,\dots,Q_k)$ is a tropical
augmentation of $M$. It is built from $W$ in time $O(n+m)$.
\end{proof}

\begin{corollary}\label{cor:MaxTM-time}
Let $M$ be a maximum matching of $G^{c}$. One can decide in time
$O(\augtime)$ whether $M$ admits a tropical augmentation. If it does,
one can compute a tropical augmentation $\mathcal{P}$ of $M$ within
the same time. Then $M\symdiff\mathcal{P}$ is a maximum matching of
$G^{c}$ that covers the colors of $M$ and one more color.
\end{corollary}

\begin{proof}
The graph $G'$ and the matching $M'$ are built from $G^{c}$ and $M$
in time $O(n+m)$. One search for an $M'$-augmenting path takes time
$O(\augtime)$. If the search fails, $M$ admits no tropical
augmentation, by \Cref{pr:MaxTM}. If it succeeds, \Cref{pr:MaxTM}(2)
gives a tropical augmentation $\mathcal{P}$ of $M$ in time $O(n+m)$.
A search reads the whole graph, so the total time is $O(\augtime)$.
The last claim is \Cref{lem:flip-aug}.
\end{proof}

\section{Minimum tropical matchings}\label{sec:min}
 The previous section gave an algorithm for computing a maximum matching that covers as many colors as possible. Here we focus instead on finding a tropical matching with as few edges as possible. The reductions used in this section decrease the matching size.
\subsection{Tropical reductions}

Our basic operation is the \emph{tropical reduction}: a chain flip that preserves the set of covered colors and decreases the matching size by one.

\begin{definition}[tropical reduction]\label{def:ktropreduc}
Let $M$ be a matching of $G^{c}$. A \emph{tropical reduction} of
$M$ of order $k\ge1$ is an $M$-flippable chain
$\mathcal{P}=(P_1,\dots,P_k)$ in which the two terminals $x_1$ and
$x'_k$ are  $M$-covered endpoints, and such that
\begin{enumerate}
\item the colors $\eta_0=\col(x_1)$ and $\eta_k=\col(x'_k)$ are
  redundant in $M$,
\item if $\eta_0=\eta_k$, then $\mult{M}{\eta_0}\ge3$.
\end{enumerate}
\end{definition}

\begin{figure}[htb]
\centering
\begin{tikzpicture}[
  vertex/.style={circle, draw, inner sep=0pt, minimum size=5.5mm,
                 font=\small},
  medge/.style={line width=2.4pt},
  oedge/.style={thin},
  pathbg/.style={line width=10pt, gray!25, line cap=round},
  cnum/.style={font=\tiny, inner sep=1pt},
  tag/.style={font=\scriptsize},
]
  \draw[pathbg] (0,0) -- (3,0);
  \draw[pathbg] (5,0) -- (6.5,0);
  \node[vertex, fill=blue!22,   label={[cnum]below:2}] (a) at (0,0) {$a$};
  \node[vertex, fill=gray!50,   label={[cnum]below:4}] (b) at (1.5,0) {$b$};
  \node[vertex, fill=yellow!55, label={[cnum]below:3}] (c) at (3,0) {$c$};
  \node[vertex, fill=yellow!55, label={[cnum]below:3}] (d) at (5,0) {$d$};
  \node[vertex, fill=white,     label={[cnum]below:1}] (e) at (6.5,0) {$e$};
  \node[vertex, fill=blue!22,   label={[cnum]below:2}] (f) at (8.5,0) {$f$};
  \node[vertex, fill=white,     label={[cnum]below:1}] (g) at (10,0) {$g$};
  \draw[medge] (a) -- (b); \draw[oedge] (b) -- (c);
  \draw[medge] (d) -- (e);
  \draw[medge] (f) -- (g);
  \draw[dotted, thick] (c) to[out=50,in=130] (d);
  \node[tag] at (4,0.8) {relay};
  \node[tag] at (0.1,0.7) {color 2 redundant};
  \node[tag] at (7.3,0.7) {color 1 redundant};
  \node[tag] at (1.7,0.6) {$P_1$};
  \node[tag] at (5.75,0.6) {$P_2$};
\end{tikzpicture}
\caption{The matching $M=\{ab,de,fg\}$ is tropical. The chain
$(P_1,P_2)$ with $P_1=a,b,c$, $P_2=d,e$ and the relay $\{c,d\}$
($c$ free, $d$ covered, color~3) is a tropical reduction: its
terminals $a$ (color 2, also carried by $f$) and $e$ (color 1, also
carried by $g$) are $M$-covered endpoints of redundant colors. After
the flip, $N=\{bc,fg\}$ is tropical with one edge less.}
\label{fig:red-ex}
\end{figure}
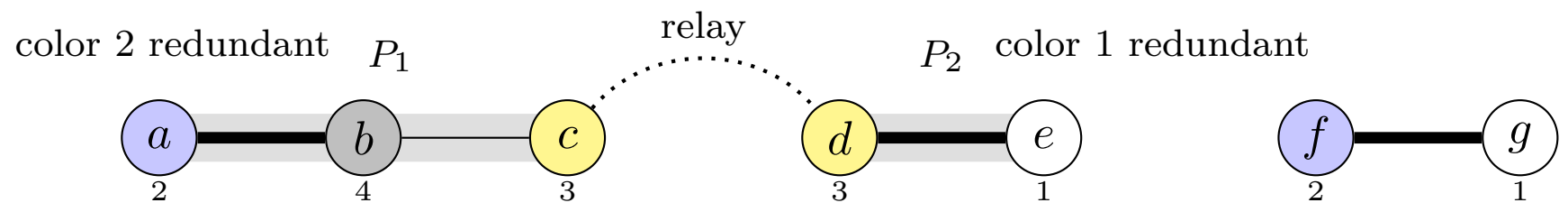

See \Cref{fig:red-ex} for an example of a tropical reduction.

\begin{lemma}[flipping a reduction]\label{lem:flip-red}
If $\mathcal{P}$ is a tropical reduction of $M$, then
$N=M\symdiff\mathcal{P}$ is a matching with $|N|=|M|-1$ and
$\colset{N}=\colset{M}$. \end{lemma}

\begin{proof}
The terminals $x_1$ and $x'_k$ are  $M$-covered endpoints, and each
relay has  one $M$-free and one $M$-covered endpoint of the same color. So
\Cref{lem:counting} gives $|N|=|M|-1$. The multiplicities of
$\eta_0$ and $\eta_k$ decrease by one, or by two for $\eta_0$ if
$\eta_0=\eta_k$, and the other multiplicities do not change. Since
$\eta_0$ and $\eta_k$ are redundant in $M$, with
$\mult{M}{\eta_0}\ge3$ if $\eta_0=\eta_k$, they stay covered by
$N$. Hence $\colset{N}=\colset{M}$.
\end{proof}

\subsection{The reduction theorem}

 As in the augmentation argument, the proof relies on a   choice of the witness matching $N$.

\begin{lemma}[choice of $N$ for reductions]\label{lem:choice-red}
Let $M$ be a matching of $G^{c}$, and assume that some matching
with the same color set as $M$ has fewer edges. Among these
matchings, let $N$ be one with $|N|$ minimum, and among those, one
with $|M\cap N|$ maximum. Let $G_{\symdiff}=(V,M\symdiff N)$. Then:
\begin{enumerate}
\item every component of $G_{\symdiff}$ with at least one edge is a
  path;  
\item if $\mathcal{Q}$ is an $N$-flippable chain made of path
  components of $G_{\symdiff}$, and if $N\symdiff\mathcal{Q}$ has
  at most $|N|$ edges, then
  $\colset{N\symdiff\mathcal{Q}}\neq\colset{M}$.
\end{enumerate}
\end{lemma}

\begin{proof}
(1) Flipping a cycle of $G_{\symdiff}$ in $N$ keeps $V(N)$, so the
size and the color set, and it brings edges of $M$ into $N$. This
contradicts the choice of $N$.

 (2) Let $N'=N\symdiff\mathcal{Q}$, a matching by
\Cref{lem:counting}  applied to $N$, and suppose that $\colset{N'}=\colset{M}$.
Since $|N'|\le|N|$, the minimality of $|N|$ gives $|N'|=|N|$.
Let $E_M=E(\mathcal{Q})\cap M$ and $E_N=E(\mathcal{Q})\cap N$.
The paths of $\mathcal{Q}$ are components of $G_{\symdiff}$, so each
edge of $\mathcal{Q}$ belongs to exactly one of $M$ and $N$.
The flip thus removes the edges of $E_N$ from $N$ and adds those of
$E_M$, and $|N'|=|N|-|E_N|+|E_M|$. As $|N'|=|N|$, we get
$|E_M|=|E_N|$: the flip adds as many edges as it removes.
Moreover $|E_M|+|E_N|\ge1$, because $\mathcal{Q}$ has at least one
edge, so $|E_M|\ge1$.
Finally, no edge of $M\cap N$ is removed, since $E_N\cap M=\emptyset$,
and the added edges are those of $E_M$. Therefore
$|M\cap N'|=|M\cap N|+|E_M|>|M\cap N|$, contrary to the choice
of $N$.\end{proof}

\begin{theorem}\label{th:red}
Let $M$ be a matching of $G^{c}$. Then $M$ admits a tropical
reduction if and only if there exists a matching $N$ of $G^{c}$
with $|N|<|M|$ and $\colset{N}=\colset{M}$.
\end{theorem}

\begin{proof}
If $\mathcal{P}$ is a tropical reduction of $M$, then
$N=M\symdiff\mathcal{P}$ has the required properties, by
\Cref{lem:flip-red}.

Conversely, suppose that such a matching exists, and let $N$ be as
in \Cref{lem:choice-red}: $\colset{N}=\colset{M}$, $|N|<|M|$, with
$|N|$ minimum and $|M\cap N|$ maximum. By \Cref{lem:choice-red}(1),
every component of $G_{\symdiff}$ with at least one edge is a path.
Let $\mathcal{Q}$ be the family of these paths. By \Cref{lem:symdiff}, each path of $\mathcal{Q}$ is
$M$-flippable and $N$-flippable. Its $M$-covered
endpoints are covered by $M$ and not by $N$. Its $M$-free
endpoints are covered by $N$ and not by $M$.
\smallskip
 
 Since $|M|>|N|$, we have
$\sum_{\eta}\mult{M}{\eta}=2|M|>2|N|=\sum_{\eta}\mult{N}{\eta}$.
So there exists a color $\eta_0$ satisfying
$\mult{M}{\eta_0}>\mult{N}{\eta_0}$.
This color is in $\colset{M}=\colset{N}$, so
$\mult{N}{\eta_0}\ge1$ and $\mult{M}{\eta_0}\ge2$.
Thus $\eta_0$ is redundant in $M$.
Moreover, $M$ covers more vertices of color $\eta_0$ than $N$, so
there exists a vertex $x_1$ of color $\eta_0$ that is covered
by $M$ and not by $N$. By \Cref{lem:symdiff}, $x_1$ is a covered
endpoint of a path $P_1\in\mathcal{Q}$. We orient $P_1$ from $x_1$
to its other endpoint $x'_1$.

Among the $M$-flippable chains $(P_1,\dots,P_k)$ of distinct paths
of $\mathcal{Q}$ that begin with $P_1$, with $P_j$ oriented from
$x_j$ to $x'_j$, choose one, $\mathcal{P}=(P_1,\dots,P_k)$, with
$k$ maximum. It exists, since $(P_1)$ is such a chain and
$\mathcal{Q}$ is finite. Set $\eta_k=\col(x'_k)$. We show that
$\mathcal{P}$ is a tropical reduction of $M$.

The two vertices of a relay have the same color, one covered by
$M$ only, the other by $N$ only. So $\mathcal{P}$ is both
$M$-flippable and $N$-flippable, and by \Cref{lem:counting},
flipping it changes the color multiplicities only at $x_1$
and $x'_k$.

\smallskip
\noindent\textit{Step 1: we prove that $x'_k$ is covered by $M$.}
Suppose, for contradiction, that $x'_k$ is covered by $N$.

\smallskip
\noindent\textit{Step 1.1:  We  prove that the
color $\eta_k$ is critical in $N$.}  Let $N'=N\symdiff\mathcal{P}$.
The flip covers $x_1$ and uncovers $x'_k$ in $N'$. By \Cref{lem:counting} applied to $N$,
$N'$ is a matching with $|N'|=|N|$, and compared to $N$ it gains
one vertex of color $\eta_0$ and loses one vertex of color
$\eta_k$, the other multiplicities being unchanged. By the choice of $N$ (\Cref{lem:choice-red}(2)),
$\colset{N'}\neq\colset{M}=\colset{N}$.
If $\eta_k=\eta_0$, all multiplicities are unchanged and
$\colset{N'}=\colset{N}$, a contradiction. So $\eta_k\neq\eta_0$.
Since $\eta_0\in\colset{N}$, the color $\eta_0$ stays in
$\colset{N'}$, and the only color that $N'$ can lose is $\eta_k$.
Hence $\mult{N}{\eta_k}=1$, as claimed.

\smallskip
\noindent\textit{Step 1.2: we extend the chain.}
Since $\eta_k\in\colset{N}=\colset{M}$, 
there exists a vertex $u$ of color $\eta_k$ covered by $M$. As
$\eta_k$ is critical in $N$, its only $N$-covered vertex is $x'_k$,
which is not covered by $M$. Hence $u\neq x'_k$, and $u$ is not
covered by $N$. By \Cref{lem:symdiff}, $u$ is an endpoint of a path
$P$ of $\mathcal{Q}$. We claim that $P$ is a new path: $P$ is none of
$P_1,\dots,P_k$. Suppose instead that $P=P_j$ for some
$j\le k$. The endpoints of $P_j$ are $x_j$ and $x'_j$, and $u$
is an endpoint of $P$. So $u$ would be $x_j$ or $x'_j$. The vertex $u$ has color $\eta_k$ and $x_1$ has color $\eta_0$:
so $u\neq x_1$. And $u\neq x'_k$, as seen above. The remaining endpoints $x'_j$ and
$x_{j+1}$, for $1\le j<k$, form the relays of $\mathcal{P}$. If
$u$ were in the relay $\{x'_j,x_{j+1}\}$, this relay would have
color $\eta_k$, and $x'_j$ would be an $N$-covered vertex of color
$\eta_k$ distinct from $x'_k$, contradicting the criticality of
$\eta_k$. This proves the claim. Set $P_{k+1}=P$, oriented from
$x_{k+1}=u$. Then $\{x'_k,x_{k+1}\}$ is a relay, so
$(P_1,\dots,P_{k+1})$ is an $M$-flippable chain of distinct paths
of $\mathcal{Q}$ beginning with $P_1$, contradicting the maximality
of $k$.

\smallskip
\noindent\textit{Step 2: we prove that $\mathcal{P}$ is a tropical reduction.}
By Step 1, $x_1$ and $x'_k$ are $M$-covered endpoints, and $\eta_0$ is
redundant in $M$. If $\eta_k$ is redundant in $M$ and either
$\eta_k\neq\eta_0$ or $\mult{M}{\eta_0}\ge3$, then $\mathcal{P}$
is a tropical reduction by \Cref{def:ktropreduc}, and $\mathcal{P}$ is a tropical reduction by \Cref{def:ktropreduc}.  It remains to rule out the other cases.

\smallskip
\noindent\textit{Step 2.1: in the remaining cases, the
$M$-covered vertices of color $\eta_k$ are among $x_1$ and
$x'_k$.}
In the remaining cases, either $\eta_k$ is critical in $M$, or
$\eta_k=\eta_0$ and $\mult{M}{\eta_0}=2$. In the first case,
$x'_k$ is the only $M$-covered vertex of color $\eta_k$. In the
second case, the only two are $x_1$ and $x'_k$. None of these
vertices is covered by $N$.

\smallskip
\noindent\textit{Step 2.2: we extend the chain.}
Since $\eta_k\in\colset{M}=\colset{N}$, there exists an
$N$-covered vertex $y$ of color $\eta_k$. Then $y$ is not
covered by $M$, so by \Cref{lem:symdiff} it is an endpoint of a
path $P$ of $\mathcal{Q}$. We claim that $P$ is a new path: $P$
is none of $P_1,\dots,P_k$. Suppose instead that $P=P_j$ for
some $j\le k$. Then $y$ would be $x_j$ or $x'_j$. Now $y$ is
covered by $N$, while $x_1$ and $x'_k$ are covered by $M$ and
not by $N$. So $y$ would be one of the remaining endpoints
$x'_j$ and $x_{j+1}$, for $1\le j<k$, which form the relays of
$\mathcal{P}$. If $y$ were in the relay $\{x'_j,x_{j+1}\}$, this
relay would have color $\eta_k$, so the other vertex of the
relay would be an $M$-covered vertex of color $\eta_k$. But the
$M$-covered vertices of color $\eta_k$ are among $x_1$ and
$x'_k$, and the vertices $x'_j$ and $x_{j+1}$, for $1\le j<k$,
are all distinct from $x_1$ and $x'_k$: the paths are disjoint,
and the two endpoints of a path are distinct. This proves the
claim. Set $P_{k+1}=P$, oriented from $x_{k+1}=y$. Then
$\{x'_k,x_{k+1}\}$ is a relay, and $(P_1,\dots,P_{k+1})$
contradicts the maximality of $k$ as in Step 1. The remaining
cases are impossible: $\mathcal{P}$ is a tropical reduction
of $M$.

\end{proof}

\begin{corollary}\label{cor:red-trop}
Let $M$ be a tropical matching of $G^{c}$. Then $M$ admits a
tropical reduction if and only if some tropical matching of
$G^{c}$ has fewer edges than $M$.
\end{corollary}

\begin{proof}
A matching is tropical if and only if its color set is $\Cols$.
Apply \Cref{th:red}.
\end{proof}

\subsection{Computing a minimum tropical matching:  Algorithm~\algmin}

We now turn \Cref{th:red} into an algorithm: start from the
matching returned by Algorithm~\algmax, and reduce it as long
as possible. Each reduction comes from one augmenting-path
search in an auxiliary graph $H$, the relay graph. We first
state the algorithm and its running time, using this search as
a black box. The relay graph is built in \Cref{sec:find-red}
below.

\begin{algorithm}[htb]
\caption{\algmin\ (maximum colorful matching with the fewest edges)}\label{alg:MinTM}
\begin{algorithmic}[1]
\State $M \gets$ the matching returned by Algorithm~\algmax\
  (\Cref{alg:MaxTM}). 
\While{$M$ admits a tropical reduction $\mathcal{P}$
  (\Cref{pr:MinTM} below)}
  \State $M \gets M\symdiff\mathcal{P}$.
\EndWhile
\State \Return $M$.
\end{algorithmic}
\end{algorithm}

\begin{theorem}\label{th:MinTM2}
Let $M$ be the matching returned by Algorithm~\algmin\
(\Cref{alg:MinTM}). If $M$ is tropical, then $M$ is a minimum
tropical matching. Otherwise $G^{c}$ has no tropical matching.
Algorithm~\algmin\ thus solves \mintmp\ in time
$O(\matchtime+n\,\augtime)=O\bigl(n\,m\,\alpha(m,n)\bigr)$.
\end{theorem}

\begin{proof}
 By \Cref{th:MaxTM2}, the matching computed at line~1 is
maximum colorful, in time $O(\matchtime+|\Cols|\,\augtime)$. By
\Cref{lem:flip-red}, each iteration of the loop keeps the color
set and removes one edge. So the color set never changes, and
there are at most $n/2$ iterations. Each iteration costs
$O(\augtime)$ by \Cref{pr:MinTM}. The final matching $M$ has the
color set of the matching of line~1. If this color set is not
$\Cols$, then $\chimax<|\Cols|$ and $G^{c}$ has no tropical
matching (\Cref{th:MaxTM2}). Otherwise $M$ is tropical and
admits no tropical reduction. By \Cref{cor:red-trop}, no
tropical matching has fewer edges: $M$ is a minimum tropical
matching. Since $|\Cols|\le n$, the total time is
$O(\matchtime+n\,\augtime)$.
\end{proof}

\subsection{Finding reductions: the relay graph}\label{sec:find-red}

We encode each relay by an alternating path in an auxiliary
graph. Adding one edge at each end then turns a tropical
reduction into an augmenting path. We proceed in three steps.
We first show that the relay colors can be chosen pairwise
distinct, so that each color gadget is used at most once. We
then define the relay graph and collect four observations on
it. Finally, we prove that tropical reductions correspond to
augmenting paths of the relay graph (\Cref{pr:MinTM}).

\begin{lemma}\label{lem:distinct-relays}
If $M$ admits a tropical reduction, it admits one whose relays
have pairwise distinct colors.
\end{lemma}

\begin{proof}
Let $\mathcal{P}=(P_1,\dots,P_k)$ be a tropical reduction of $M$
of minimum order $k$. We suppose that two relays
$\{x'_i,x_{i+1}\}$ and $\{x'_j,x_{j+1}\}$, with $i<j$, have the
same color $\eta$. Let $u=x'_i$ and $v=x_{j+1}$. We distinguish
three cases, according to which of $u$ and $v$ are $M$-covered.

If exactly one of $u,v$ is $M$-covered, then $\{u,v\}$ is a
relay. We use it to join $P_i$ directly to $P_{j+1}$, deleting
$P_{i+1},\dots,P_j$. The resulting chain has the same terminals
as $\mathcal{P}$ and smaller order. It is therefore a tropical
reduction. This contradicts the choice of $\mathcal{P}$ with
$k$ minimum.

If $u,v$ are both $M$-covered, we consider the chain
$(P_1,\dots,P_i)$. Its last terminal is $u$. The vertex $v$ is
another $M$-covered vertex of color $\eta$, distinct from $u$
since the paths are disjoint.

If $u,v$ are both $M$-free, their relay partners
$x_{i+1},x'_j$ are $M$-covered. We consider the chain
$(P_1,\dots,P_j)$. Its last terminal is $x'_j$. The vertex
$x_{i+1}$ is another $M$-covered vertex of color $\eta$,
distinct from $x'_j$: either they lie on distinct paths, or
$i+1=j$ and they are the two endpoints of~$P_j$.

In both cases, the chain considered has smaller order. Its
terminals are $M$-covered. The color $\eta$ of its last
terminal is redundant in $M$. If $\eta=\col(x_1)$, then $x_1$
and the two $M$-covered vertices above are three distinct
$M$-covered vertices of color $\eta$, so $\mult{M}{\eta}\ge3$.
The chain considered is thus a tropical reduction. This
contradicts the choice of $\mathcal{P}$ with $k$ minimum.
\end{proof}

We now define the auxiliary graph. It contains $G$, one gadget
for each color, and a few extra vertices that control the two
terminals of a reduction.

\begin{definition}[relay graph]\label{def:relay-graph}
Let $M$ be a matching of $G^{c}$. The \emph{relay graph}
$H=H(G^{c},M)$, with its matching $M_H$, is defined as follows.
\begin{itemize}
\item $H$ contains $G$ as a subgraph, and $M_H$ contains $M$.
\item For every $M$-free vertex $v$, $H$ has a new vertex $r_v$
  and the edge $r_v v$, which belongs to $M_H$.
\item For every color $\eta$ occurring in $G$, $H$ has two new
  vertices $s_\eta$ and $t_\eta$ and the edge $s_\eta t_\eta$,
  which belongs to $M_H$. Moreover, $s_\eta$ is adjacent to
  every $M$-covered vertex of color $\eta$, and $t_\eta$ is
  adjacent to $r_v$ for every $M$-free vertex $v$ of color
  $\eta$.
\item For every color $\eta$ with $\mult{M}{\eta}=2$, $H$ has a
  new vertex $b_\eta$, adjacent to the two $M$-covered vertices
  of color $\eta$. Finally, $H$ has two new vertices $b^{(1)}$
  and $b^{(2)}$, both adjacent to every $M$-covered vertex whose
  color has multiplicity at least $3$ in $M$. The vertices
  $b_\eta$, $b^{(1)}$, and $b^{(2)}$ are the \emph{sink
  vertices}.
\end{itemize}
The vertices of $G$ are the \emph{original vertices} of $H$, and
the other vertices are the \emph{added vertices}.
\end{definition}

The relay graph is not much larger than $G$. There are at most
$n$ colors occurring in $G$, and each original vertex receives
only a constant number of new incident edges. Hence $H$ has
$O(n)$ vertices and $O(m+n)$ edges.

We now describe the vertices that $M_H$ leaves free. All added
vertices are distinct, and the matching $M_H$ consists exactly
of the edges of $M$, the edges $r_v v$ for the $M$-free
vertices $v$, and the edges $s_\eta t_\eta$ for the colors
$\eta$ occurring in $G$. Every original vertex is covered once,
either by $M$ or by an edge $r_v v$, and each $s_\eta$ is
matched to $t_\eta$. This gives the following observation.

\begin{observation}\label{obs:free}
The $M_H$-free vertices of $H$ are exactly the sink vertices.
\end{observation}

An augmenting path of $H$ therefore starts and ends at sink
vertices. The next observation explains the role of these
vertices: they encode the two conditions of
\Cref{def:ktropreduc}. See \Cref{fig:sink-gadgets}.

\begin{observation}\label{obs:sinks}
Two distinct $M$-covered vertices of $G$ have two distinct sink
neighbors in $H$ if and only if their colors are redundant in
$M$ and, when their colors coincide, their multiplicity is at
least three.
\end{observation}

\begin{proof}
Let $u$ and $v$ be two distinct $M$-covered vertices. The sink
neighbors of $u$ are: none if $\col(u)$ is critical in $M$; the
single vertex $b_{\col(u)}$ if $\mult{M}{\col(u)}=2$; both
$b^{(1)}$ and $b^{(2)}$ if $\mult{M}{\col(u)}\ge3$. The same
holds for $v$. If $\col(u)$ or $\col(v)$ is critical, one of
them has no sink neighbor. If $\col(u)=\col(v)$ with
multiplicity $2$, both have the single sink neighbor
$b_{\col(u)}$. In all the other cases, $u$ and $v$ have two
distinct sink neighbors.
\end{proof}

We now turn to the relays. A relay $\{u,v\}$ of color $\eta$,
with $u$ $M$-covered and $v$ $M$-free, is represented by the
path $u,s_\eta,t_\eta,r_v,v$, or its reverse. Its edges are
successively outside, inside, outside, and inside $M_H$. At
$u$, it joins the edge $M(u)$ of a flippable path. At $v$, it
joins an edge of $G$ outside $M$. Thus it preserves alternation
at both ends.  
The next observation is the converse. An augmenting path can go
from $G$ to $G$ through added vertices only along such a gadget
path.

\begin{observation}\label{obs:gadget}
Let $W$ be an $M_H$-augmenting path of $H$. Let $u$ and $v$ be
two vertices of $G$ on $W$ such that, between them, $W$ visits
only added vertices, at least one. Then the portion of $W$ from
$u$ to $v$ is $u,s_\eta,t_\eta,r_v,v$, or its reverse, and
$\{u,v\}$ is a relay of color $\eta$.
\end{observation}

\begin{proof}
The vertices of $W$ strictly between $u$ and $v$ are added
vertices. They are internal vertices of $W$, so they are
$M_H$-covered: by \Cref{obs:free}, they are not sink vertices.
The vertex $s_\eta$ is adjacent to $t_\eta$ and to $M$-covered
vertices of color $\eta$. The vertex $t_\eta$ is adjacent to
$s_\eta$ and to the vertices $r_v$ with $\col(v)=\eta$. The
vertex $r_v$ is adjacent to $t_{\col(v)}$ and to $v$. So, up to
exchanging $u$ and $v$, the portion is $u,s_\eta,t_\eta,r_v,v$,
with $u$ $M$-covered, $v$ $M$-free, and
$\col(u)=\col(v)=\eta$: the pair $\{u,v\}$ is a relay.
\end{proof}

It remains to look at the parts of an augmenting path that lie
in $G$. They will be the paths of our chain. A \emph{maximal
subpath of $W$ in $G$} is a subpath of $W$ whose vertices all
lie in $G$ and that cannot be extended.

\begin{observation}\label{obs:pieces}
Let $W$ be an $M_H$-augmenting path of $H$. Every maximal
subpath of $W$ in $G$ has at least one edge. Moreover, each of
its $M$-covered endpoints has its $M$-edge on the subpath.
\end{observation}

\begin{proof}
Let $P$ be a maximal subpath of $W$ in $G$, and let $u$ be an
endpoint of $P$. It is an internal vertex of $W$, since the
endpoints of $W$ are sink vertices (\Cref{obs:free}). If $u$
is $M$-covered, its edges of $W$ outside $G$ are not in $M_H$,
so its $M_H$-edge on $W$ is $M(u)$, an edge of $G$: it lies on
$P$. If $u$ is $M$-free, its only edge of $H$ outside $G$ is
$ur_u$, so its other edge of $W$ lies in $G$ and on $P$. In
both cases, $P$ has an edge at $u$.
\end{proof}

We can now prove the main result of this subsection.
\Cref{obs:free,obs:sinks} handle the two terminals of the
chain. \Cref{obs:gadget,obs:pieces} handle its relays and its
paths.

\begin{proposition}\label{pr:MinTM}
Let $M$ be a matching of $G^{c}$. Then $M$ admits a tropical
reduction if and only if $H$ has an $M_H$-augmenting path. A
tropical reduction can be recovered from any such path in time
$O(n+m)$. Hence a tropical reduction of $M$, if any, can be
found in time $O(\augtime)$.
\end{proposition}

\begin{proof}
\noindent\textit{Step 1: from a tropical reduction to an
augmenting path.} We suppose first that $M$ admits a tropical reduction. By
\Cref{lem:distinct-relays}, we choose one,
$\mathcal{P}=(P_1,\dots,P_k)$, with pairwise distinct relay
colors. We replace each relay by the path of its gadget, and we
add a sink neighbor at each terminal, choosing two distinct
sink vertices, as \Cref{def:ktropreduc} and \Cref{obs:sinks}
allow. This gives a walk $W$ in $H$. It is a path: the paths
$P_i$ are vertex-disjoint, the relay colors are pairwise
distinct, each $r_v$ belongs to a distinct endpoint, and the
two sink vertices are distinct. It alternates, since the gadget
paths preserve alternation. Its endpoints are $M_H$-free by
\Cref{obs:free}. So $W$ is an $M_H$-augmenting path.

\smallskip
\noindent\textit{Step 2: from an augmenting path to a tropical
reduction.} Conversely, let $W$ be an $M_H$-augmenting path. By
\Cref{obs:free}, its endpoints are two distinct sink vertices.
Let $P_1,\dots,P_k$ be the maximal subpaths of $W$ in $G$, in
their order along $W$. They are vertex-disjoint, and their
edges alternate between $M$ and $E\setminus M$, since
$M_H\cap E=M$. By \Cref{obs:gadget}, consecutive subpaths are
joined by relays. By \Cref{obs:pieces}, each subpath has at
least one edge, and each $M$-covered endpoint has its $M$-edge
on its subpath. By \Cref{obs:flippable}, each $P_i$ is
$M$-flippable. So $(P_1,\dots,P_k)$ is an $M$-flippable chain.
Its terminals are adjacent in $W$ to two distinct sink
vertices, so by \Cref{obs:sinks} their colors satisfy
\Cref{def:ktropreduc}. The chain is a tropical reduction
of $M$.

\smallskip
\noindent\textit{Step 3: the   time complexity.} Constructing $H$ and $M_H$ takes time $O(n+m)$, and extracting
the paths $P_i$ takes one scan of $W$. Since $H$ has $O(n)$
vertices and $O(m+n)$ edges, the search runs in time
$O(\augtime)$ (\Cref{sec:complexity}).
\end{proof}

\section{Minimum maximum colorful matchings and reconfiguration}\label{sec:colmax}

 This section gathers two results on maximum colorful matchings. \Cref{sec:exchange} shows that Algorithm~\algmin\ returns one with the fewest edges; the tool is an exchange lemma. \Cref{sec:reconf} shows that flips connect all maximum matchings that are maximum colorful.

\subsection{The exchange lemma}\label{sec:exchange}
When no tropical matching exists, we can focus on a matching that covers the maximum number $\chimax$ of colors and, subject to that, has the fewest edges. The algorithm~\algmax\ returns a matching covering $\chimax$ colors (\Cref{th:MaxTM2}), and \Cref{th:red} gives a way to reduce its size without changing its color set. Since maximum colorful matchings may cover different sets of colors, we first prove the exchange lemma below.

\begin{lemma}[exchange]\label{lem:exchange}
Let $M$ and $N$ be two color-maximum matchings of $G^{c}$. There
is a color-maximum matching $N'$ with $|N'|=|N|$ and
$\colset{N'}=\colset{M}$.
\end{lemma}

\begin{proof}
Among the  colorful matchings with exactly $|N|$ edges
(there is at least one, $N$ itself), we choose $N'$ with $|M\cap N'|$
maximum. Suppose that $\colset{N'}\neq\colset{M}$. Both sets have
$\chimax$ elements, so there exist a color $\eta$   in $\colset{M}$ and not
in $\colset{N'}$.

\smallskip
 
By construction, there is a vertex $w$ of color $\eta$ covered by $M$ and free
for $N'$. In $G_{\symdiff}=(V,M\symdiff N')$, $w$ has
exactly one edge, its $M$-edge. So $w$ is the endpoint of a path
component $P$ of $G_{\symdiff}$. By \Cref{lem:symdiff}, the path
$P$ is $N'$-flippable, and $w$ is one of its $N'$-free endpoints. Let $w'$ be the other endpoint of $P$, and let
$L=N'\symdiff E(P)$: a matching, since $P$ is $N'$-flippable.  We consider two
cases.

\smallskip
\noindent\textit{Case 1: $w'$ is free for $N'$.}
Then both endpoints of $P$ are free for $N'$: the path $P$ is
$N'$-augmenting. The matching $L$ covers $V(N')$, plus $w$ and
$w'$. So it covers every color of $N'$, and also $\eta$: at least
$\chimax+1$ colors. This case is impossible.

\smallskip
\noindent\textit{Case 2: $w'$ is covered by $N'$.}
Then $P$ has one $N'$-free endpoint and one $N'$-covered endpoint, so
$|L|=|N'|=|N|$. The matching $L$ covers $w$, it does not cover
$w'$, and it covers the other vertices of $V(N')$. So $\colset{L}$
contains color $\eta$, and it contains every color of $N'$, except maybe
$\col(w')$. If no color disappeared, $L$ would cover at least
$\chimax+1$ colors. So the color $\col(w')$ disappears, and
$\colset{L}$ has exactly $\chimax$ elements: the matching $L$ is
maximum colorful, with $|N|$ edges. It is a candidate for the choice
of $N'$.

We compare $M\cap L$ and $M\cap N'$. The edges of $M\cap N'$ are
not in $G_{\symdiff}$, so they are not on $P$, and they stay in
$L$. The edges of $E(P)\cap M$ enter $L$, and there is at least
one, the $M$-edge of $w$. So $|M\cap L|>|M\cap N'|$. This
contradicts the choice of $N'$.

\smallskip
Both cases are impossible. So $\colset{N'}=\colset{M}$.
\end{proof}

Algorithm~\algmin\ (\Cref{alg:MinTM}) already does the compute the color-maximum matching with the fewest edges. Its first line computes a maximum colorful matching, and its loop removes edges without changing the color set. The question is whether the final matching has the fewest edges among all maximum colorful matchings, and this is where the exchange lemma enters.

\begin{corollary}\label{cor:colmax-min}
The matching returned by Algorithm~\algmin\ (\Cref{alg:MinTM}) is a maximum colorful matching with the minimum number of edges among the maximum colorful matchings. It is computed in time
$O(\matchtime+n\,\augtime)\modif{=O\bigl(n\,m\,\alpha(m,n)\bigr)}$.
\end{corollary}

\begin{proof}
The running time, and the fact that the final matching is maximum colorful, are in the proof of \Cref{th:MinTM2}.

Let $M$ be the final matching: it admits no tropical reduction.
Let $N$ be a maximum colorful matching with $|N|<|M|$. By
\Cref{lem:exchange}, some matching $N'$ satisfies $|N'|=|N|<|M|$
and $\colset{N'}=\colset{M}$. By \Cref{th:red}, the matching $M$
admits a tropical reduction: a contradiction. So no maximum colorful
matching has fewer edges than $M$.
\end{proof}

\subsection{Reconfiguration}\label{sec:reconf}

Reconfiguration asks whether any two solutions of a problem are
connected by a sequence of simple moves~\cite{Ito11,Nishimura18}.
For maximum matchings in an uncolored graph, the answer is
immediate: it suffices to flip the components of $M\symdiff N$
one at a time. Each cycle or path flip preserves maximum
cardinality. In the colored setting, a single path flip may
uncover a critical color. Flippable chains let us coordinate
several path flips and avoid this problem.

\begin{figure}[htb]
\centering
\begin{tikzpicture}[
  vertex/.style={circle, draw, inner sep=0pt, minimum size=5.5mm,
                 font=\small},
  medge/.style={line width=2.4pt},
  nedge/.style={line width=2.4pt, densely dashed},
  oedge/.style={thin},
  cnum/.style={font=\tiny, inner sep=1pt},
  tag/.style={font=\scriptsize},
  rowlabel/.style={anchor=east, font=\small},
]
  \begin{scope}
    \node[vertex, fill=gray!50,  label={[cnum]below:$\eta_1$}] (u1) at (0,0) {$u_1$};
    \node[vertex, fill=white,    label={[cnum]below:$\eta_2$}] (v1) at (1.5,0) {$v_1$};
    \node[vertex, fill=blue!22,  label={[cnum]below:$\eta_3$}] (w1) at (3,0) {$w_1$};
    \node[vertex, fill=blue!22,  label={[cnum]below:$\eta_3$}] (u2) at (5,0) {$u_2$};
    \node[vertex, fill=white,    label={[cnum]below:$\eta_2$}] (v2) at (6.5,0) {$v_2$};
    \node[vertex, fill=gray!50,  label={[cnum]below:$\eta_1$}] (w2) at (8,0) {$w_2$};
    \draw[nedge] (u1) -- (v1); \draw[medge] (v1) -- (w1);
    \draw[nedge] (u2) -- (v2); \draw[medge] (v2) -- (w2);
    \draw[dotted, thick] (w1) to[out=50,in=130] (u2);
    \node[tag] at (4,0.85) {relay};
    \node[rowlabel] at (-0.8,0) {(a)};
  \end{scope}
  \begin{scope}[yshift=-2.2cm]
    \node[vertex, fill=gray!50,  label={[cnum]below:$\eta_1$}] (u1) at (0,0) {$u_1$};
    \node[vertex, fill=white,    label={[cnum]below:$\eta_2$}] (v1) at (1.5,0) {$v_1$};
    \node[vertex, fill=blue!22,  label={[cnum]below:$\eta_3$}] (w1) at (3,0) {$w_1$};
    \node[vertex, fill=blue!22,  label={[cnum]below:$\eta_3$}] (u2) at (5,0) {$u_2$};
    \node[vertex, fill=white,    label={[cnum]below:$\eta_2$}] (v2) at (6.5,0) {$v_2$};
    \node[vertex, fill=gray!50,  label={[cnum]below:$\eta_1$}] (w2) at (8,0) {$w_2$};
    \draw[medge] (u1) -- (v1); \draw[oedge] (v1) -- (w1);
    \draw[oedge] (u2) -- (v2); \draw[medge] (v2) -- (w2);
    \node[tag] at (3,0.75) {the color $\eta_3$ is missing};
    \node[rowlabel] at (-0.8,0) {(b)};
  \end{scope}
  \begin{scope}[yshift=-4.4cm]
    \node[vertex, fill=gray!50,  label={[cnum]below:$\eta_1$}] (u1) at (0,0) {$u_1$};
    \node[vertex, fill=white,    label={[cnum]below:$\eta_2$}] (v1) at (1.5,0) {$v_1$};
    \node[vertex, fill=blue!22,  label={[cnum]below:$\eta_3$}] (w1) at (3,0) {$w_1$};
    \node[vertex, fill=blue!22,  label={[cnum]below:$\eta_3$}] (u2) at (5,0) {$u_2$};
    \node[vertex, fill=white,    label={[cnum]below:$\eta_2$}] (v2) at (6.5,0) {$v_2$};
    \node[vertex, fill=gray!50,  label={[cnum]below:$\eta_1$}] (w2) at (8,0) {$w_2$};
    \draw[medge] (u1) -- (v1); \draw[oedge] (v1) -- (w1);
    \draw[medge] (u2) -- (v2); \draw[oedge] (v2) -- (w2);
    \node[tag] at (4,0.75) {tropical again: this is $N$};
    \node[rowlabel] at (-0.8,0) {(c)};
  \end{scope}
\end{tikzpicture}
\caption{(a) The matchings $M$ (thick) and $N$ (dashed); the two
paths of $G_{\symdiff}$ and their relay $\{w_1,u_2\}$ of color
$\eta_3$. (b) Flipping the path $u_1,v_1,w_1$ alone: the vertex
$w_1$ is uncovered, the color $\eta_3$ is missing, the matching is
not a vertex of $\mathcal{R}(G^{c})$. (c) Flipping the chain of
the two paths: the matching is $N$, tropical. One  move  connects
$M$ and $N$.}
\label{fig:reconf-ex}
\end{figure}
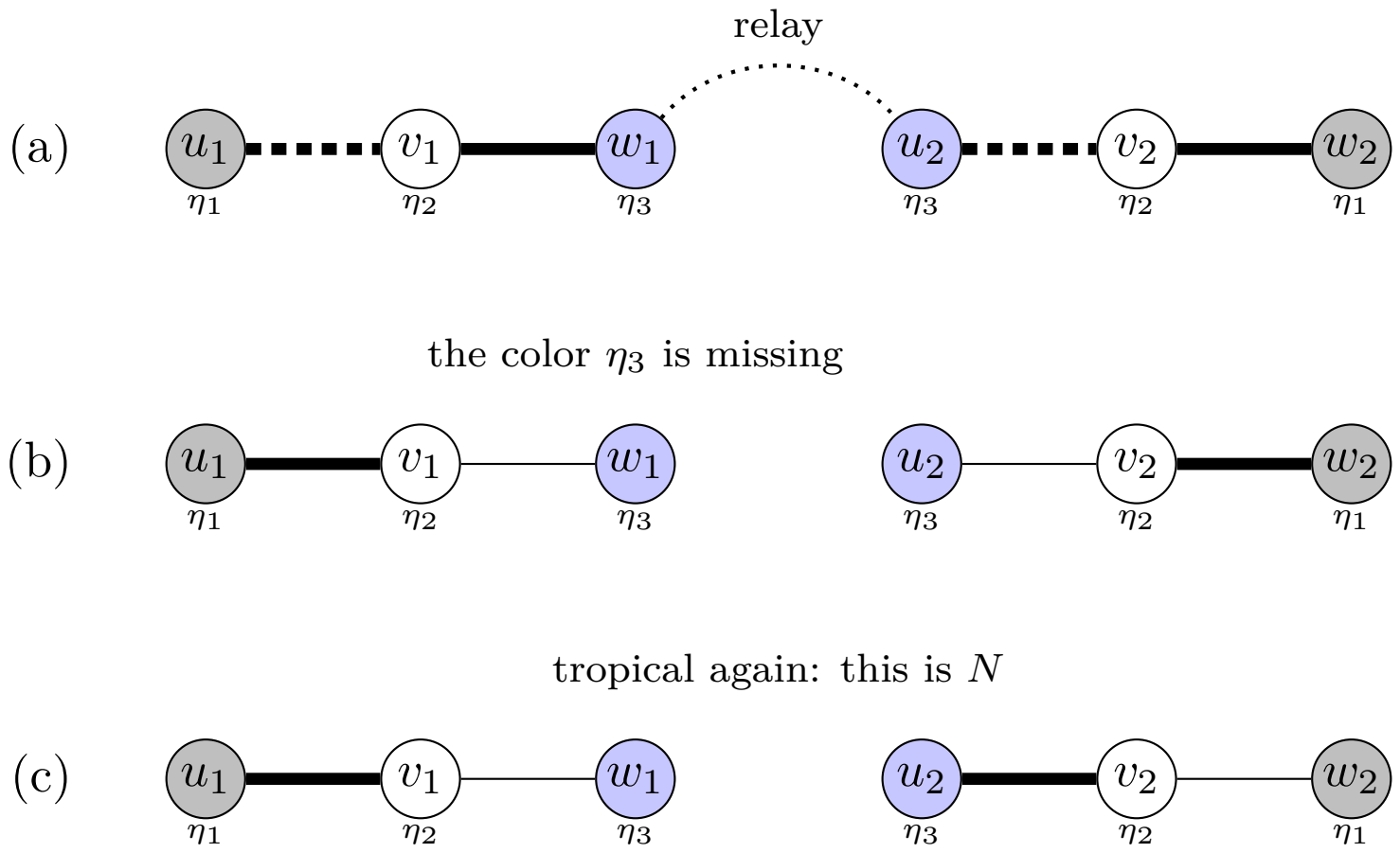

\begin{example}\label{ex:reconf}
We consider two disjoint paths $u_1,v_1,w_1$ and
$u_2,v_2,w_2$, with colors $\eta_1,\eta_2,\eta_3$ and
$\eta_3,\eta_2,\eta_1$, and the tropical maximum matchings
$M=\{v_1w_1,v_2w_2\}$ and $N=\{u_1v_1,u_2v_2\}$ of
\Cref{fig:reconf-ex}. Flipping one path of $G_{\symdiff}$ alone
uncovers $w_1$ or $w_2$, the only covered vertex of its color:
the result is no longer tropical. The chain of the two paths,
linked by the relay $\{w_1,u_2\}$ of color $\eta_3$, transforms
$M$ into $N$ in one move: the multiplicities of $\eta_1$ and
$\eta_3$ do not change.
\end{example}

\begin{definition}[move, reconfiguration graph]\label{def:reconf}
Let $M$ be a maximum matching of $G^{c}$ that is maximum
colorful. A \emph{move} transforms $M$ into
$M\symdiff\mathcal{P}$, for an $M$-flippable chain
$\mathcal{P}$, or into $M\symdiff E(C)$, for a cycle $C$ whose
edges alternate between $M$ and $E\setminus M$. The
\emph{reconfiguration graph} $\mathcal{R}(G^{c})$ has one
vertex for each maximum matching of $G^{c}$ that is maximum
colorful. Two vertices $M$ and $N$ are adjacent if a move
transforms $M$ into $N$.
\end{definition}

We prove that $\mathcal{R}(G^{c})$ is connected
(\Cref{th:reconf}). The idea is the following. If $M$ and $N$
are two distinct vertices of $\mathcal{R}(G^{c})$, then one move
transforms $M$ into a vertex $M'$ with
$M'\symdiff N\subsetneq M\symdiff N$. Since $|M\symdiff N|$ is a
nonnegative integer, repeating such moves leads from $M$ to $N$.
The move from $M$ to $M'$ is the flip of one component of
$G_{\symdiff}=(V,M\symdiff N)$, a cycle or a path, or of a chain
of such paths. The difficulty is to choose the flip so that $M'$
remains maximum colorful.

Throughout, $M$ and $N$ are two distinct vertices of
$\mathcal{R}(G^{c})$, and $G_{\symdiff}=(V,M\symdiff N)$. Both
are maximum matchings, so \Cref{lem:two-max} applies: every path
of $G_{\symdiff}$ has a start and an end, and we orient it from
its start to its end.

We first check that such flips bring $M$ closer to $N$. Let $M'$
be obtained from $M$ by flipping a cycle of $G_{\symdiff}$, or a
chain as in \Cref{lem:two-max}(3). A cycle flip keeps the set of
covered vertices, and a chain flip keeps the size of $M$. So
$M'$ is a maximum matching. On the flipped edges, $M'$ agrees
with $N$, and it agrees with $M$ elsewhere. So $M'\symdiff N$ is
$M\symdiff N$ without the flipped edges, and there is at least
one. This gives the following observation.

\begin{observation}\label{obs:reconf-closer}
Let $M'$ be obtained from $M$ by flipping a cycle of
$G_{\symdiff}$, or a chain as in \Cref{lem:two-max}(3). Then
$M'$ is a maximum matching with
$M'\symdiff N\subsetneq M\symdiff N$. It is a neighbor of $M$ in
$\mathcal{R}(G^{c})$.

\end{observation}

We now treat two cases, according to whether
$\colset{M}\neq\colset{N}$ (\Cref{lem:reconf-diff}) or
$\colset{M}=\colset{N}$ (\Cref{lem:reconf-eq}).

\begin{lemma}[the color sets differ]\label{lem:reconf-diff}
If $\colset{M}\neq\colset{N}$, then $G_{\symdiff}$
has a path $P$ such that $M'=M\symdiff E(P)$ is a neighbor of
$M$ in $\mathcal{R}(G^{c})$. Moreover,
$M'\symdiff N\subsetneq M\symdiff N$.
\end{lemma}

\begin{proof}
Both color sets have $\chimax$ elements and differ. So there
exists a color $\alpha\in\colset{N}\setminus\colset{M}$. Let $u$
be a vertex of color $\alpha$ covered by $N$. Since
$\alpha\notin\colset{M}$, the vertex $u$ is free for $M$. By
\Cref{lem:two-max}(2), it is the start of a path $P_1$ of
$G_{\symdiff}$. Let $\beta$ be the color of the end of $P_1$.
This end is covered by $M$, so $\beta\in\colset{M}$ and
$\beta\neq\alpha$.

Let $M'=M\symdiff E(P_1)$. By \Cref{lem:counting}, the flip
covers the start of $P_1$, uncovers its end, and changes nothing
else. So $M'$ covers $\alpha$ and every color of $\colset{M}$
other than $\beta$. If $\beta$ were redundant in $M$, then $M'$
would also cover $\beta$. It would cover the $\chimax+1$ colors
of $\colset{M}\cup\{\alpha\}$, which is impossible. So $\beta$
is critical in $M$, and
$\colset{M'}=(\colset{M}\setminus\{\beta\})\cup\{\alpha\}$ has
$\chimax$ elements. By \Cref{obs:reconf-closer}, $M'$ is a
neighbor of $M$ with $M'\symdiff N\subsetneq M\symdiff N$.
\end{proof}

\begin{lemma}[the color sets are equal]\label{lem:reconf-eq}
If $\colset{M}=\colset{N}$, then one move, the flip
of one cycle of $G_{\symdiff}$ or of one $M$-flippable chain
made of paths of $G_{\symdiff}$, transforms $M$ into a vertex
$M'$ of $\mathcal{R}(G^{c})$ with
$M'\symdiff N\subsetneq M\symdiff N$.
\end{lemma}

\begin{proof}
\noindent\textit{Case 1: $G_{\symdiff}$ has a cycle $C$.}
Let $M'=M\symdiff E(C)$. A cycle flip keeps the set of covered
vertices, so $\colset{M'}=\colset{M}$. By
\Cref{obs:reconf-closer}, $M'$ is a neighbor of $M$ with
$M'\symdiff N\subsetneq M\symdiff N$.

\smallskip
\noindent\textit{Case 2: $G_{\symdiff}$ has no cycle.}
Since $M\neq N$, the graph $G_{\symdiff}$ has a path $P_1$. Let
$\alpha$ be the color of its start. We consider the sequences
$(P_1,\dots,P_k)$ of distinct paths of $G_{\symdiff}$ beginning
with $P_1$, in which the end of $P_i$ and the start of $P_{i+1}$
have the same color for $1\le i<k$. We choose one with $k$
maximum. It exists, since $(P_1)$ is such a sequence and
$G_{\symdiff}$ has finitely many paths. Let $\beta$ be the color
of the end of $P_k$.

\smallskip
\noindent\textit{Step 1: we prove that $\beta$ is redundant in
$M$ or $\beta=\alpha$.}
We suppose, for contradiction, that $\beta$ is critical in $M$
and $\beta\neq\alpha$. Then the end of $P_k$ is the only
$M$-covered vertex of color $\beta$. Since
$\beta\in\colset{M}=\colset{N}$, there exists a vertex $y$ of
color $\beta$ covered by $N$. It is not the end of $P_k$, which
is free for $N$. So $y$ is free for $M$. By
\Cref{lem:two-max}(2), $y$ is the start of a path $P$ of
$G_{\symdiff}$.

We claim that $P$ is a new path: $P$ is none of
$P_1,\dots,P_k$. We suppose instead that $P=P_j$ for some
$j\le k$. Then $y$ is the start of $P_j$. We have $j\neq1$,
since the start of $P_1$ has color $\alpha\neq\beta$. And if
$2\le j\le k$, the end of $P_{j-1}$ has color $\beta$ and is
covered by $M$. It is distinct from the end of $P_k$, since
$j-1<k$. This is a second $M$-covered vertex of color $\beta$:
a contradiction. This proves the claim. So $(P_1,\dots,P_k,P)$
is a longer sequence of the same kind. This contradicts the
maximality of $k$.

\smallskip
\noindent\textit{Step 2: we prove that the flip is a move.}
By \Cref{lem:two-max}(3), $(P_1,\dots,P_k)$ is an $M$-flippable
chain. Let $M'$ be the matching obtained by flipping it. By
\Cref{lem:counting}, the multiplicities change only at the two
terminals: the flip adds one covered vertex of color $\alpha$
and removes one of color $\beta$. The start of $P_1$ is covered
by $N$, so $\alpha\in\colset{N}=\colset{M}$: no color is gained.
If $\beta=\alpha$, no multiplicity changes. Otherwise, $\beta$
is redundant in $M$ by Step~1, so it stays covered. In both
cases, $\colset{M'}=\colset{M}$. By \Cref{obs:reconf-closer},
$M'$ is a neighbor of $M$ with
$M'\symdiff N\subsetneq M\symdiff N$.
\end{proof}

\begin{theorem}\label{th:reconf}
The graph $\mathcal{R}(G^{c})$ is connected.
\end{theorem}

\begin{proof}
Let $M$ and $N$ be two vertices of $\mathcal{R}(G^{c})$. We
prove by induction on $|M\symdiff N|$ that $\mathcal{R}(G^{c})$
has a path from $M$ to $N$. If $M\symdiff N=\emptyset$, then
$M=N$. Otherwise, \Cref{lem:reconf-diff} or
\Cref{lem:reconf-eq} gives a neighbor $M'$ of $M$ with
$|M'\symdiff N|<|M\symdiff N|$. By induction,
$\mathcal{R}(G^{c})$ has a path from $M'$ to $N$, and the edge
$MM'$ extends it to $M$.
\end{proof}
  
\section{Matchings of maximum color weight}\label{sec:matroid}

We assign to each color $\eta$ a weight $\omega(\eta)\ge0$, and
write $\omega(S)=\sum_{\eta\in S}\omega(\eta)$ for a set of colors
$S$. The \emph{color weight} of a matching $M$ is
$\omega(\colset{M})$, and we want to compute a matching of maximum
color weight. With unit weights, this is a maximum colorful matching,
computed by Algorithm~\algmax\ in \Cref{sec:max}. For general weights, the
problem remains polynomial: \Cref{th:weighted2} below computes a
matching of maximum color weight that is also a maximum matching,
in time $O(|\Cols|\,\matchtime+|\Cols|^{2}\,\augtime)$.

 Our tool is a matroid, and it comes from \Cref{sec:max}. The algorithm~\algmax\ computes a maximum colorful matching (\Cref{th:MaxTM2}). The augmentation theorem behind it (\Cref{th:MaxTM}) gives more: the sets of colors that a matching can cover form a matroid on $\Cols$ (\Cref{sec:color-matroid}). The greedy algorithm then finds such a set of maximum weight, and  Algorithm~\algmax\ serves as its independence oracle (\Cref{sec:greedy}). Finally, a classical description of this matroid gives its rank function, hence a min-max formula for $\chimax$ and a variant of Hall's theorem: a tropical matching exists if and only if, for every set $J$ of colors, some matching covers at least $|J|$ vertices whose colors lie in $J$ (\Cref{sec:hall}).

\subsection{Coverable sets of colors}\label{sec:coverable}

We say that a set of colors $S\subseteq\Cols$ is \emph{coverable}
if some matching $M$ of $G^{c}$ satisfies $S\subseteq\colset{M}$.
The empty set of the colors is coverable, and every subset of a coverable set is coverable, by the same matching. We test coverability with Algorithm~\algmax, on the modified instance $G^{c}_{S'}$ defined below.

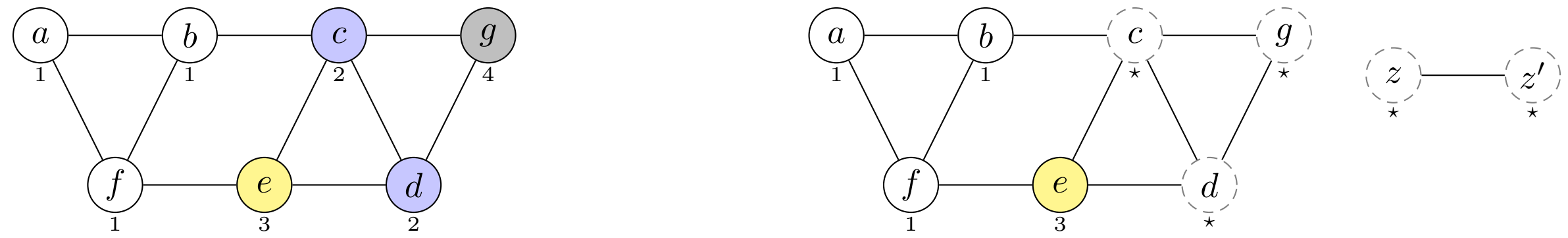
\begin{figure}[htb]
\centering
\begin{tikzpicture}[
  vertex/.style={circle, draw, inner sep=0pt, minimum size=5.5mm,
                 font=\small},
  star/.style={circle, draw=gray, densely dashed, inner sep=0pt,
               minimum size=5.5mm, font=\small, fill=white},
  medge/.style={line width=2.4pt},
  oedge/.style={thin},
  cnum/.style={font=\tiny, inner sep=1pt},
  tag/.style={font=\scriptsize},
]
  \begin{scope}
    \node[vertex, fill=white,     label={[cnum]below:1}] (a) at (0,0) {$a$};
    \node[vertex, fill=white,     label={[cnum]below:1}] (b) at (1.5,0) {$b$};
    \node[vertex, fill=blue!22,   label={[cnum]below:2}] (c) at (3,0) {$c$};
    \node[vertex, fill=gray!50,   label={[cnum]below:4}] (g) at (4.5,0) {$g$};
    \node[vertex, fill=white,     label={[cnum]below:1}] (f) at (0.75,-1.5) {$f$};
    \node[vertex, fill=yellow!55, label={[cnum]below:3}] (e) at (2.25,-1.5) {$e$};
    \node[vertex, fill=blue!22,   label={[cnum]below:2}] (d) at (3.75,-1.5) {$d$};
    \draw[oedge] (a)--(b); \draw[oedge] (b)--(c); \draw[oedge] (c)--(g);
    \draw[oedge] (a)--(f); \draw[oedge] (b)--(f); \draw[oedge] (c)--(e);
    \draw[oedge] (f)--(e); \draw[oedge] (d)--(e); \draw[oedge] (c)--(d);
    \draw[oedge] (g)--(d);
    \node[tag] at (2.25,-2.6) {(a) a graph $G^{c}$, and the test of $S\cup\{\eta\}=\{1,3\}$};
  \end{scope}
  \begin{scope}[xshift=8cm]
    \node[vertex, fill=white,     label={[cnum]below:1}] (a) at (0,0) {$a$};
    \node[vertex, fill=white,     label={[cnum]below:1}] (b) at (1.5,0) {$b$};
    \node[star, label={[cnum]below:$\star$}] (c) at (3,0) {$c$};
    \node[star, label={[cnum]below:$\star$}] (g) at (4.5,0) {$g$};
    \node[vertex, fill=white,     label={[cnum]below:1}] (f) at (0.75,-1.5) {$f$};
    \node[vertex, fill=yellow!55, label={[cnum]below:3}] (e) at (2.25,-1.5) {$e$};
    \node[star, label={[cnum]below:$\star$}] (d) at (3.75,-1.5) {$d$};
    \node[star, label={[cnum]below:$\star$}] (z) at (5.6,-0.4) {$z$};
    \node[star, label={[cnum]below:$\star$}] (zp) at (7,-0.4) {$z'$};
    \draw[oedge] (a)--(b); \draw[oedge] (b)--(c); \draw[oedge] (c)--(g);
    \draw[oedge] (a)--(f); \draw[oedge] (b)--(f); \draw[oedge] (c)--(e);
    \draw[oedge] (f)--(e); \draw[oedge] (d)--(e); \draw[oedge] (c)--(d);
    \draw[oedge] (g)--(d);
    \draw[oedge] (z)--(zp);
    \node[tag] at (3.5,-2.6) {(b) the graph $G^{c}_{\{1,3\}}$};
  \end{scope}
\end{tikzpicture}
\caption{The coverability test by Algorithm~\algmax. The set $\{1,3\}$ is
coverable in (a) if and only if the graph $G^{c}_{\{1,3\}}$ in
(b), with
colors $1$, $3$ and $\star$, has a tropical matching. In (b), the
vertices with a color outside $\{1,3\}$ are recolored $\star$
(dashed), and the edge $zz'$ makes the color $\star$ free of
charge.}
\label{fig:oracle}
\end{figure}

\begin{definition}[modified instance $G^{c}_{S'}$]\label{def:modified}
Let $S'\subseteq\Cols$ be a set of colors, and let $z$, $z'$ be
two new vertices. The \emph{modified instance} $G^{c}_{S'}$ is the
vertex-colored graph built from $G^{c}$ as follows:
\begin{enumerate}
\item its vertices are $V\cup\{z,z'\}$ and its edges are
  $E\cup\{zz'\}$;
\item a vertex of $V$ whose color in $G^{c}$ is in $S'$ keeps
  this color;
\item the other vertices, including $z$ and $z'$, get a single new
  color $\star$.
\end{enumerate}
The colors of $G^{c}_{S'}$ are $S'\cup\{\star\}$. See
\Cref{fig:oracle}.
\end{definition}

\begin{lemma}\label{lem:modified}
A set of colors $S'$ is coverable in $G^{c}$ if and only if
$G^{c}_{S'}$ has a tropical matching $K$. In this
case, $K\cap E(G)$ is a matching of $G$ that covers every color
of $S'$.
\end{lemma}

\begin{proof}
Let $M$ be a matching of $G$ that covers every color of $S'$.
Then $M\cup\{zz'\}$ covers $S'$ and $\star$: it is a tropical
matching of $G^{c}_{S'}$. Conversely, let $K$ be a tropical
matching of $G^{c}_{S'}$. Each color of $S'$ is
covered by $K$, and only by edges of $G$, since $zz'$ covers only
the color $\star$. So $K\cap E(G)$ covers every color of $S'$.
\end{proof}

\begin{remark}
The coverability test is performed by Algorithm~\algmax.
We construct the modified instance $G^{c}_{S'}$ of
\Cref{def:modified} and run \algmax\ on it. If \algmax\ returns a
tropical matching of $G^{c}_{S'}$, that is, a matching covering
all colors of $S'\cup\{\star\}$, then $S'$ is coverable; otherwise
it is not (\Cref{lem:modified}). By \Cref{th:MaxTM2}, the test
takes time $O(\matchtime+(|S'|+1)\,\augtime)$.\end{remark}

\begin{lemma}[color exchange]\label{lem:color-exchange}
Let $A$ and $B$ be two coverable sets of colors with $|B| > |A|$. Then $A\cup\{\eta\}$ is coverable for some color $\eta\in B\setminus A$.
\end{lemma}

\begin{proof}
Let $H^{c}=G^{c}_{A\cup B}$ be the modified instance of
\Cref{def:modified}, and let $H$ be its underlying graph. The
colors of $H^{c}$ are $A\cup B\cup\{\star\}$. Since $A$ is
coverable, there exists a matching $K$ of $G$ that covers every
color of $A$. Then $K\cup\{zz'\}$ is a matching of $H$ that covers
$A\cup\{\star\}$. By \Cref{lem:berge-ext}, applied to $H$, $H$ 
admits   a maximum matching  $M$  that covers every vertex
covered by $K\cup\{zz'\}$, hence every color of $A\cup\{\star\}$.
If $M$ covers a color $\eta\in B\setminus A$, then $M\cap E(G)$
covers $A\cup\{\eta\}$, since the edge $zz'$ covers only the color
$\star$. In this case $A\cup\{\eta\}$ is coverable, as required.

Now, we  assume that $\colset{M}=A\cup\{\star\}$. Likewise, since $B$ is coverable, \Cref{lem:berge-ext} gives a
maximum matching $N$ of $H$ that covers $B\cup\{\star\}$. So
$|\colset{N}|\ge|B|+1>|A|+1=|\colset{M}|$. By \Cref{th:MaxTM},
applied to $H^{c}$, the matching $M$ admits a tropical
augmentation $\mathcal{P}$. By \Cref{lem:flip-aug},
$\colset{M\symdiff\mathcal{P}}=\colset{M}\cup\{\eta\}$ for a color
$\eta\notin\colset{M}$. This color lies in $A\cup B\cup\{\star\}$
and not in $A\cup\{\star\}$, so $\eta\in B\setminus A$. Finally,
$(M\symdiff\mathcal{P})\cap E(G)$ is a matching of $G$ that covers
$A\cup\{\eta\}$,  because $zz'$ covers only the color $\star$.
So $A\cup\{\eta\}$ is coverable.
\end{proof}

\Cref{lem:color-exchange} has a   consequence: the coverable
sets of colors form a matroid. We now make this precise.

\subsection{The color matroid}\label{sec:color-matroid}

\begin{definition}[matroid]\label{def:matroid}
A \emph{matroid} on a finite set $Y$ is a family of subsets of $Y$,
called \emph{independent}, such that
\begin{enumerate}
\item the empty set is independent,
\item every subset of an independent set is independent,
\item if $A$ and $B$ are independent and $|A|<|B|$, then
  $A\cup\{y\}$ is independent for some $y\in B\setminus A$.
\end{enumerate}
The maximal independent sets are the \emph{bases}. By item 3, they
all have the same size, the \emph{rank} of the matroid. The rank
$r(U)$ of a set $U\subseteq Y$ is the size of the largest
independent set inside $U$.
\end{definition}

The coverable sets of colors satisfy the three items. The empty
set is covered by the empty matching: item 1 holds. A matching
that covers a set of colors also covers all its subsets: item 2
holds. Item 3 is
exactly \Cref{lem:color-exchange}, with $A$ and $B$ the two
coverable sets.

\begin{theorem}[color matroid]\label{th:color-matroid}
The coverable sets of colors form a matroid on $\Cols$. Its rank is
$\chimax$. Its bases are the color sets $\colset{M}$ of the
maximum colorful matchings $M$.
\end{theorem}

\begin{proof}
We checked above that the coverable sets
of colors satisfy the three items of \Cref{def:matroid}: they
form a matroid on $\Cols$.

Let $S$ be a coverable set, covered by a colorful matching $M$.
Then $S$ has at most $|\colset{M}|\le\chimax$ elements. And the
color set of a maximum colorful matching has $\chimax$ elements.
So the rank is $\chimax$.

A base is a coverable set $B$ with $\chimax$ elements. Let $M$
be a colorful matching covering $B$. Then
$|\colset{M}|\le\chimax=|B|$ forces $\colset{M}=B$: the matching
$M$ is maximum colorful. Conversely, the color set of a maximum
colorful matching is coverable and has $\chimax$ elements: it is
a base.

\end{proof}

The color matroid is a classical object. It is the matroid induced from the matching matroid of $G$~\cite{EdmondsFulkerson65} by the color classes, in the sense of Rado~\cite{Rado42} and Perfect~\cite{Perfect69}. We use this description in \Cref{sec:hall}, where it gives the rank function of the color matroid.

\subsection{The greedy algorithm}\label{sec:greedy}

Item 3, the exchange axiom, is the usefull property of matroids: it
is what makes the greedy algorithm work on them, as we will see
in this remaining of this section.

\begin{theorem}[greedy, Rado~\cite{Rado57};
Edmonds~\cite{Edmonds71}]\label{th:greedy}
Let $\omega$ be a weight $\omega(y)\ge0$ on each element of a
matroid on $Y$, and write $\omega(S)=\sum_{y\in S}\omega(y)$.
Examine the elements of $Y$ by decreasing weight, starting from
$S=\emptyset$, and add each element $y$ to $S$ when $S\cup\{y\}$
is independent. The final set $S$ is a base, and
$\omega(S)\ge\omega(I)$ for every independent set $I$.
\end{theorem}

In our setting, the weights only order the colors; they
play no role in the matching computations. Since the coverable
color sets form a matroid (\Cref{th:color-matroid}), the greedy
theorem (\Cref{th:greedy}) applies. The algorithm only needs the
coverability test of \Cref{lem:modified}.

\begin{algorithm}[htb]
\caption{\algweight\ (maximum color weight)}\label{alg:weighted}
\begin{algorithmic}[1]
\State $S \gets \emptyset$; \ $M_0 \gets \emptyset$.
\For{each color $\eta$, by decreasing weight $\omega(\eta)$}
  \State Run Algorithm~\algmax\ on the graph $G^{c}_{S\cup\{\eta\}}$
    (\Cref{def:modified}).
  \If{it returns a tropical matching $K$}
    \State $S \gets S\cup\{\eta\}$; \ $M_0 \gets K\cap E(G)$.
  \EndIf
\EndFor
\State Extend $M_0$ to a maximum matching $M$ of $G$ by augmenting
  paths.
\State \Return $M$.
\end{algorithmic}
\end{algorithm}

\begin{theorem}\label{th:weighted2}
\algweight\ (\Cref{alg:weighted}) computes a matching of maximum
size and maximum color weight, in time
$O(|\Cols|\,\matchtime+|\Cols|^{2}\,\augtime)\modif{=O\bigl(|\Cols|\,\sqrt{n}\,m+|\Cols|^{2}\,m\,\alpha(m,n)\bigr)}$.
\end{theorem}

\begin{proof}
The loop adds $\eta$ to $S$ when $S\cup\{\eta\}$ is coverable:
this is the test of \Cref{lem:modified}. So the loop is the
greedy algorithm of the color matroid. By \Cref{th:greedy}, the
final set $S$ is a base $B$, and $\omega(B)\ge\omega(I)$ for
every coverable set $I$.

Consider the last successful test. The tested set is the final
set $B$. By \Cref{lem:modified}, the matching $M_0$ found there
covers $B$: $\colset{M_0}\supseteq B$. Moreover
$|\colset{M_0}|\le\chimax=|B|$, so $\colset{M_0}=B$. The
extension keeps the covered vertices covered, as in
\Cref{lem:berge-ext}. So $M$ is a maximum matching with
$\colset{M}=B$. Now take any matching $N$ of $G^{c}$. Its color
set is coverable, so $\omega(\colset{N})\le\omega(B)$. The
matching $M$ has the maximum color weight.

There are $|\Cols|$ iterations. Each one is a call to
Algorithm~\algmax, in time $O(\matchtime+|\Cols|\,\augtime)$ by
\Cref{th:MaxTM2}. The final extension is one more matching
computation. The total time is
$O(|\Cols|\,\matchtime+|\Cols|^{2}\,\augtime)$.
\end{proof}

A good choice of weights can also force colors to be covered:
give a large weight to some colors, a small one to the others.
This gives the following.

\begin{corollary}[mandatory colors]\label{cor:mandatory}
Let $P$ be a coverable set of colors. A matching of maximum size, and of maximum color weight among the matchings that cover every color of $P$, can be computed in time $O(|\Cols|\,\matchtime+|\Cols|^{2}\,\augtime)$.
\end{corollary}

\begin{proof}
We run \algweight\ with $S=P$ at the start. The initial matching
$M_0$ is a matching of $G$ that covers $P$, given by one call to
Algorithm~\algmax\ on $G^{c}_{P}$ (\Cref{lem:modified}). The
loop examines only the colors outside $P$.

Consider the sets $A\subseteq\Cols\setminus P$ such that
$P\cup A$ is coverable. They form a matroid on
$\Cols\setminus P$. Items 1 and 2 are clear. For item 3, apply
\Cref{lem:color-exchange} to $P\cup A$ and $P\cup A'$: it gives
a color $\eta\in A'\setminus A$ with $P\cup A\cup\{\eta\}$
coverable. This matroid is the contraction of the color matroid
by $P$.

The loop is the greedy algorithm of this matroid. So the final
set $S$ has the maximum weight among the sets $A$ with $P\cup A$
coverable, and $P\cup S$ is a base of the color matroid. The
proof of \Cref{th:weighted2} then applies with $B=P\cup S$.
\end{proof}

\section{Conclusion}\label{sec:conclusion}

Our approach is based on a single idea: Berge's augmenting paths~\cite{B57}, adapted to colors. A tropical augmentation is a chain of flippable paths. Two consecutive paths are linked by a relay, a covered vertex and a free vertex of the same color. Flipping the chain keeps the size of the matching and adds a color. A tropical reduction is a chain of the same kind. Flipping it keeps the colors and removes an edge. A maximum matching covers the most colors if and only if it admits no tropical augmentation (\Cref{th:MaxTM}). A matching has the fewest edges among the matchings with its color set if and only if it admits no tropical reduction (\Cref{th:red}). Both are found by an ordinary augmenting-path search in an auxiliary graph: the quotient graph $G'$ for augmentations, the relay graph $H$ for reductions. The algorithms of \Cref{tab:summary} follow. 

The same chains give structural results. Flips of chains and of alternating cycles connect all maximum matchings that are maximum colorful. In other words, the reconfiguration graph $\mathcal{R}(G^{c})$ is connected (\Cref{th:reconf}). The augmentation theorem has a second consequence. The sets of colors that a matching can cover form a matroid (\Cref{th:color-matroid}). On this matroid, the greedy algorithm computes a matching of maximum color weight, with Algorithm~\algmax\ as its oracle. 

\begin{table}[htb]
\centering\small
\setlength{\tabcolsep}{6pt}
\renewcommand{\arraystretch}{1.25}
\begin{tabular}{@{}>{\raggedright\arraybackslash}p{5.6cm}
                  >{\raggedright\arraybackslash}p{3.8cm}
                  >{\raggedright\arraybackslash}p{4.8cm}@{}}
\toprule
Problem & Algorithm & Complexity \\
\midrule
Computing a maximum matching with the most colors (\maxtmp)
  & Algorithm~\algmax\ (\Cref{alg:MaxTM})
  & polynomial, $O(\mathcal{T}+|\Cols|\,\mathcal{A})$ (\Cref{th:MaxTM2}) \\
\corrige{Computing a tropical matching with the fewest edges (\mintmp)}
  & Algorithm~\algmin\ (\Cref{alg:MinTM})
  & polynomial, $O(\mathcal{T}+n\,\mathcal{A})$ (\Cref{th:MinTM2}, \Cref{cor:colmax-min}) \\
Computing a matching of maximum color weight
  & \algweight\ (\Cref{alg:weighted})
  & polynomial, $O(|\Cols|\,\mathcal{T}+|\Cols|^{2}\,\mathcal{A})$ (\Cref{th:weighted2}) \\
\bottomrule
\end{tabular}
\caption{\corrige{The three algorithms of the paper. Here $\mathcal{T}$ is
the time of a maximum matching computation and $\mathcal{A}$ the time
of an augmenting-path search (\Cref{sec:complexity}), with
$\mathcal{T}=O(\sqrt{n}\,m)$~\cite{MicaliVazirani80} and
$\mathcal{A}=O(m\,\alpha(m,n))$~\cite{Gabow76,Tarjan83}.}}
\label{tab:summary}
\end{table}

In this paper, one covered vertex is enough to cover a color. A natural extension gives each color $\eta$ a demand $k_\eta\ge1$. The matching must then cover at least $k_\eta$ vertices of color $\eta$. A color is missing, critical, or redundant when its multiplicity in $M$ is below, equal to, or above its demand. Flips, relays, chains, and the counting lemma extend to demands without modifications. \Cref{th:MaxTM,th:red} do not extend in their current form.  A color may be covered and still fall short of its demand. Chains must be allowed to use such colors, and the auxiliary graphs change. We leave the extension of these two theorems to demands open. Similar demands appear in recent work on matchings and covers with group requirements~\cite{BandyapadhyayBanikBhore23,DoronAradKulikShachnai23,DabasKhullerRivkin25}.

\bibliographystyle{plain}
\bibliography{redaction-article}
\appendix

\section{A variant of Hall's theorem}\label{sec:hall}

This appendix gives a consequence of the color matroid. It
explains why a tropical matching may fail to exist.

Algorithm~\algmax\ decides whether $G^{c}$ has a tropical
matching (\Cref{th:MaxTM2}). When the answer is yes, the
matching itself is a proof: anyone can check it. When the answer
is no, the algorithm gives no such proof. To be convinced, one
has to run it again. The theorem of this appendix gives a proof
for the answer no. It is the analogue, for colors, of Hall's
theorem for bipartite graphs (see~\cite{LovaszPlummer86}).
There, a set of vertices with too few neighbors certifies that
no matching covers all these vertices.

Here, the certificate is a set of colors $J$ such that no
matching covers $|J|$ vertices whose colors lie in $J$. Such a
set proves that no tropical matching exists: a tropical matching
would cover at least one vertex of each color of $J$. Checking
the certificate does not compute a tropical matching. It
computes the largest number of vertices with a color in $J$
that one matching can cover. This is one weighted matching
computation, with weights $0$, $1$ or $2$ on the edges.

The theorem also gives a min-max formula for $\chimax$: every
set $J$ bounds the number of colors that a matching can cover,
and some $J$ reaches the bound.

The proof uses the description of the color matroid given at
the end of \Cref{sec:color-matroid}. We need two classical
theorems. The first one gives the matroid we start from.

\begin{theorem}[matching matroid, Edmonds and
Fulkerson~\cite{EdmondsFulkerson65}]\label{th:matching-matroid}
The sets of vertices $U\subseteq V$ that some matching of $G$
covers entirely form a matroid on $V$. Its rank function is
\[
r_G(U)=\max\{\,|V(M)\cap U| : M \text{ matching of } G\,\}.
\]
\end{theorem}

The second theorem builds a new matroid from a given one. It
needs one definition.

\begin{definition}[independent transversal]\label{def:transversal}
Let $\mathcal{M}$ be a matroid on $Y$, and let $(Y_x)_{x\in X}$
be a family of subsets of $Y$, indexed by a finite set $X$. An
\emph{independent transversal} of a set $S\subseteq X$ is an
injection $\varphi\colon S\to Y$ such that
$\varphi(x)\in Y_x$ for every $x\in S$, and such that the set
$\varphi(S)$ is independent in $\mathcal{M}$.
\end{definition}

In words: we choose one representative in $Y_x$ for each element
$x$ of $S$. The representatives are all distinct, and together
they form an independent set.

\begin{theorem}[induced matroid, Rado~\cite{Rado42};
Perfect~\cite{Perfect69}]\label{th:induced}
With the notation of \Cref{def:transversal}, let $r$ be the rank
function of $\mathcal{M}$. The subsets of $X$ that have an
independent transversal form a matroid on $X$. Its rank function
is
\[
\rho(T)=\min_{J\subseteq T}
\Bigl(\,r\Bigl(\bigcup_{x\in J}Y_x\Bigr)+|T\setminus J|\,\Bigr)
\qquad\text{for every } T\subseteq X .
\]
\end{theorem}

With $T=X$, the formula says: the set $X$ itself has an
independent transversal if and only if
$r\bigl(\bigcup_{x\in J}Y_x\bigr)\ge|J|$ for every $J\subseteq X$.
This is the Hall condition, with a rank in place of a
cardinality.

We now come back to colors. For a set of colors
$J\subseteq\Cols$, let $V_J$ be the set of vertices whose color
is in $J$. Then $r_G(V_J)$ is the largest number of vertices of
$V_J$ that a single matching of $G$ covers, as in
\Cref{th:matching-matroid}.

\begin{theorem}[Hall condition for colors]\label{th:hall}
The vertex-colored graph $G^{c}$ has a tropical matching if and
only if, for every set of colors $J$, some matching of $G$
covers at least $|J|$ vertices of $V_J$. More generally,
\[
\chimax=\min_{J\subseteq\Cols}
\bigl(\,r_G(V_J)+|\Cols\setminus J|\,\bigr).
\]
\end{theorem}

\begin{proof}
\noindent\textit{Step 1: we give an upper bound.}
Let $M$ be a matching and $J$ a set of colors. At most
$|\Cols\setminus J|$ colors of $M$ lie outside $J$. Each color
of $M$ that lies in $J$ has a covered vertex in $V_J$, and these
vertices are distinct. So there are at most $r_G(V_J)$ such
colors. Hence $|\colset{M}|\le r_G(V_J)+|\Cols\setminus J|$ for
every $J$, and $\chimax$ is at most the minimum. In particular,
a tropical matching covers at least $|J|$ vertices of $V_J$ for
every $J$: one for each color of $J$.

\smallskip
\noindent\textit{Step 2: we prove that the color matroid is an
induced matroid.}
We apply \Cref{th:induced} with $X=\Cols$, $Y=V$, with
$Y_\eta=\cls{\eta}$ for every color $\eta$, and with
$\mathcal{M}$ the matching matroid of $G$
(\Cref{th:matching-matroid}). We claim that a set of colors has
an independent transversal if and only if it is coverable.

We suppose first that a set $S$ of colors has an independent
transversal $\varphi$. The set $\varphi(S)$ is independent in
the matching matroid, so some matching $M$ covers it entirely.
For each $\eta\in S$, the matching covers the vertex
$\varphi(\eta)$, of color $\eta$. So $S\subseteq\colset{M}$: the
set $S$ is coverable.

Conversely, we suppose that $S$ is coverable. Let $M$ be a
matching with $S\subseteq\colset{M}$. For each $\eta\in S$, we
choose a vertex $\varphi(\eta)$ of color $\eta$ covered by $M$.
The color classes are disjoint, so $\varphi$ is injective, and
$\varphi(S)\subseteq V(M)$ is independent: $\varphi$ is an
independent transversal of $S$. This proves the claim. So the
color matroid is the induced matroid of \Cref{th:induced}, with
$\bigcup_{\eta\in J}Y_\eta=V_J$.

\smallskip
\noindent\textit{Step 3: we give the formula.}
The rank of the color matroid is $\chimax$
(\Cref{th:color-matroid}). The rank formula of
\Cref{th:induced}, with $T=\Cols$, gives
$\chimax=\min_{J\subseteq\Cols}\bigl(r_G(V_J)+|\Cols\setminus J|\bigr)$.
Finally, a tropical matching exists if and only if
$\chimax=|\Cols|$. Since
$r_G(V_J)+|\Cols\setminus J|=|\Cols|+r_G(V_J)-|J|$, this holds
if and only if $r_G(V_J)\ge|J|$ for every $J$.
\end{proof}

\begin{figure}[htb]
\centering
\begin{tikzpicture}[
  vertex/.style={circle, draw, inner sep=0pt, minimum size=5.5mm,
                 font=\small},
  medge/.style={line width=2.4pt},
  oedge/.style={thin},
  cnum/.style={font=\tiny, inner sep=1pt},
  tag/.style={font=\scriptsize},
]
  \begin{scope}
    \node[vertex, fill=white,     label={[cnum]below:1}] (a) at (0,0) {$a$};
    \node[vertex, fill=blue!22,   label={[cnum]below:2}] (b) at (-1.2,-1.3) {$b$};
    \node[vertex, fill=yellow!55, label={[cnum]below:3}] (c) at (1.2,-1.3) {$c$};
    \draw[oedge] (a) -- (b); \draw[oedge] (a) -- (c);
    \node[tag, align=center] at (0,-2.5)
      {(a) for $J=\{2,3\}$, every matching\\ covers at most one vertex of $V_J$};
  \end{scope}
  \begin{scope}[xshift=6.5cm, yshift=-0.65cm]
    \node[vertex, fill=blue!22,   label={[cnum]below:2}] (b) at (0,0) {$b$};
    \node[vertex, fill=white,     label={[cnum]below:1}] (a) at (1.6,0) {$a$};
    \node[vertex, fill=yellow!55, label={[cnum]below:3}] (c) at (3.2,0) {$c$};
    \node[vertex, fill=white,     label={[cnum]below:1}] (d) at (4.8,0) {$d$};
    \draw[medge] (b) -- (a); \draw[oedge] (a) -- (c); \draw[medge] (c) -- (d);
    \node[tag, align=center] at (2.4,-1.85)
      {(b) every set $J$ passes the test;\\ the matching $\{ba,cd\}$ is tropical};
  \end{scope}
\end{tikzpicture}
\caption{The Hall condition of \Cref{th:hall}. (a) A star: the set
$J=\{2,3\}$ fails, and no tropical matching exists. (b) A path:
for $J=\{2,3\}$, the matching $\{ba,cd\}$ covers $b$ and $c$, two
vertices of $V_J$. Every set $J$ passes, and $\{ba,cd\}$ is a
tropical matching.}
\label{fig:hall}
\end{figure}
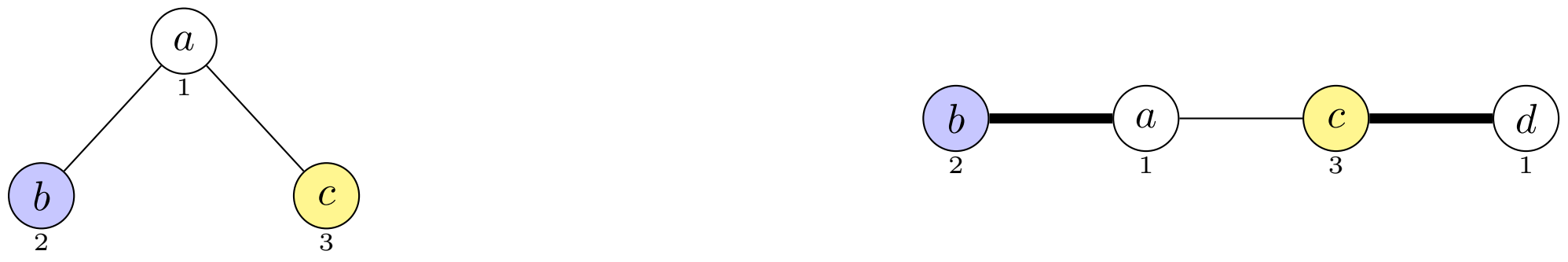

The Hall condition is the special case $\chimax=|\Cols|$. At
first, it may seem too weak: it only asks for $|J|$ covered
vertices in $V_J$, and they could all have the same color.
Rado's theorem shows that asking this for every set $J$ is
nevertheless enough. \Cref{fig:hall} shows two examples.

\end{document}